%% file: COSILab_ArXiv.tex
\documentclass{article} 
\usepackage{iclr2027_preprint,times}

\usepackage{hyperref}
\usepackage{url}
\usepackage{xspace}
\usepackage{graphicx}
\usepackage{amsmath, amssymb}
\usepackage{dsfont}
\usepackage{microtype}      
\usepackage{xcolor}         
\usepackage{tabularx}
\usepackage{booktabs}       
\usepackage{mdframed}
\usepackage{subcaption}
\usepackage{listings}
\usepackage{threeparttable}
\usepackage{caption}
\usepackage{array}
\usepackage[utf8]{inputenc} 
\usepackage[T1]{fontenc}    

\usepackage{amsfonts}       
\usepackage{nicefrac}       
\usepackage{pifont}
\definecolor{tablecheck}{HTML}{5FAF82}
\definecolor{tablecross}{HTML}{D96F63}
\newcommand{\cmark}{\textcolor{tablecheck}{\ensuremath{\checkmark}}}
\newcommand{\xmark}{\textcolor{tablecross}{\ensuremath{\times}}}

\usepackage[most]{tcolorbox}
\usepackage{xcolor}

\definecolor{promptbg}{HTML}{F5F5F0}
\definecolor{promptblue}{HTML}{2C5F7C}

\newtcolorbox{promptbox}{
    enhanced,
    breakable,
    colback=promptbg,
    colframe=promptblue,
    leftrule=2.5pt,
    rightrule=0pt,
    toprule=0pt,
    bottomrule=0pt
}

\newcommand{\promptsection}[1]{%
    \par\medskip
    {\large\bfseries #1}
    \par\smallskip
}

\newcommand{\promptsubsection}[1]{%
    \par\smallskip
    {\bfseries #1}
    \par\smallskip
}

\title{COSI-Lab:\\ Conference Living Lab for Modeling Multi-Perspective Multimodal Social Intention}

\author{%
Zonghuan Li$^{1,*}$ \quad
Litian Li$^{1,*}$ \quad
Arthur Mercier$^{1,*}$ \quad \\
Gara Dorta$^{1}$ \quad 
Balint Dioszegi$^{2}$ \quad
Jose Morales-Vargas$^{3}$ \quad
Chenxu Hao $^{1}$ \quad 
Ivan Kondyurin$^{1, **}$ \quad  \\
Vanessa Begemann$^{4}$ \quad 
Nale Lehmann-Willenbrock$^{4}$ \quad 
Bernd Dudzik$^{1}$ \quad 
Saunaq Chakrabarty$^{1}$ \quad \\
Sotiris Vacanas$^{1}$ \quad 
Laura Cabrera-Quirós$^{3}$ \quad 
Anne L.J. ter Wal$^{5}$ \quad 
Vitaliy Popov$^{6}$ \quad \\
Jorge Castro-Godínez$^{3}$ \quad
Chirag Raman$^{1,\ddagger}$ \quad
Stephanie Tan$^{1,\dagger}$ \quad
Hayley Hung$^{1,\dagger}$ \quad \\
\\
$^{1}$Delft University of Technology, 
\texttt{\{z.li-25, l.li-11, a.mercier, g.dorta, } \\
\texttt{c.hao-1, s.vacanas, c.a.raman, b.dudzik, s.tan-1, h.hung\}@tudelft.nl}; \\
$^{2}$University of Greenwich, \texttt{b.dioszegi@greenwich.ac.uk}; \\
$^{3}$Instituto Tecnológico de Costa Rica, \texttt{josfemova@estudiantec.cr, } \\
\texttt{\{lcabrera, jocastro\}@itcr.ac.cr}; 
$^{4}$University of Hamburg, \\
\texttt{\{vanessa.begemann, nale.lehmann-willenbrock\}@uni-hamburg.de}; \\
$^{5}$Imperial College London, \texttt{a.terwal@imperial.ac.uk};\\
$^{6}$University of Michigan Ann Arbor, \texttt{vipopov@umich.edu}; \\
$^{*}$Equal contribution \quad
$^{\ddagger}$Co-lead \quad
$^{\dagger}$Corresponding author \quad 
$^{**}$Work performed while affiliated \\
with Delft University of Technology. Current email: \texttt{ivan.kondyurin@gmail.com}.
}

\iclrpreprintcopy

\begin{document}

\maketitle

\begin{abstract}
COSI-Lab presents a multimodal multi-sensor dataset of a real interdisciplinary scientific workshop with 32 academics at an international  conference. It captures ecologically valid social interactions in a weakly scripted setting consisting of two 30-minute mingling sessions with real professional and social consequences. We argue that intelligent systems could better handle subjective perceptions by modeling their multiplicity not as label noise but as an explainable perspective-driven reasoning process. We focus on the Apparent Intent Inference (AII) problem as determined by ex-situ observers and conceptualize intentions to be independent of the future. We contribute a novel annotation process for AII that accounts for a perceiver's own interpretative tendencies, quantitative and qualitative analyses of intent narratives with respect to diversity, plausibility and grounding, benchmark tasks for AII and relevant contextual factors, speech quality audio for all participants as well as privacy-preserving multi-modal environmental and wearable sensor data, enabling lexical and nonverbal behavior analysis; and the coupling of self-reported participant goals (30 minute to 3 hour) with annotated AII (seconds).

\end{abstract}
\vspace{-2pt}
\section{Introduction}
\vspace{-2pt}
Human intention-aware systems are crucial for understanding future outcomes \citep{belardinelli2023gazebased,humantrajpredict2020,rasouli2019pie,jingpsi}. We argue that to understand future outcomes, and those that could influence the affective state of an observed subject, such systems need to reason about the evidence for an intention perception \citep{jingpsi} rather than just detecting them \citep{rasouli2019pie,Zhang_2022}. And do this in in-the-wild ecologically valid settings.

Following \citet{bratman1987intention}, intent forms part of future action planning coupled with the belief the action can be carried out. We focus on interpreting behavior to study \emph{overt intentional action} \citep{mele1992springs} as a proximal phenomenon (intentions related to immediate actions). We name this the Apparent Intent Inference (AII) problem. Embracing the subjectivity of AII, we take a perspectivist stance \citep{Cabitza2023, Dudzik2019context} by coupling annotators' trait, demographic data, and cognitive style  with their annotations. Finally, we also consider distal intentions/goals which relate to the future but are not necessarily situated in the immediate situation. 

Following script theory from \citet{schank_abelson_1977}, our AII problem focuses on a weakly scripted networking mingling events \citep{raman2022conflab} where the rules of play are not stated apriori(e.g. when to talk to whom, on what topic, or why are emergent). This contrasts with more popular strongly scripted settings, externally pre-determined rules e.g. rules of the road \citep{rasouli2019pie}, situational comedy shows \citep{zhang2024mintrec}.

\begin{figure*}[t]
    \centering
    \includegraphics[width=\textwidth]{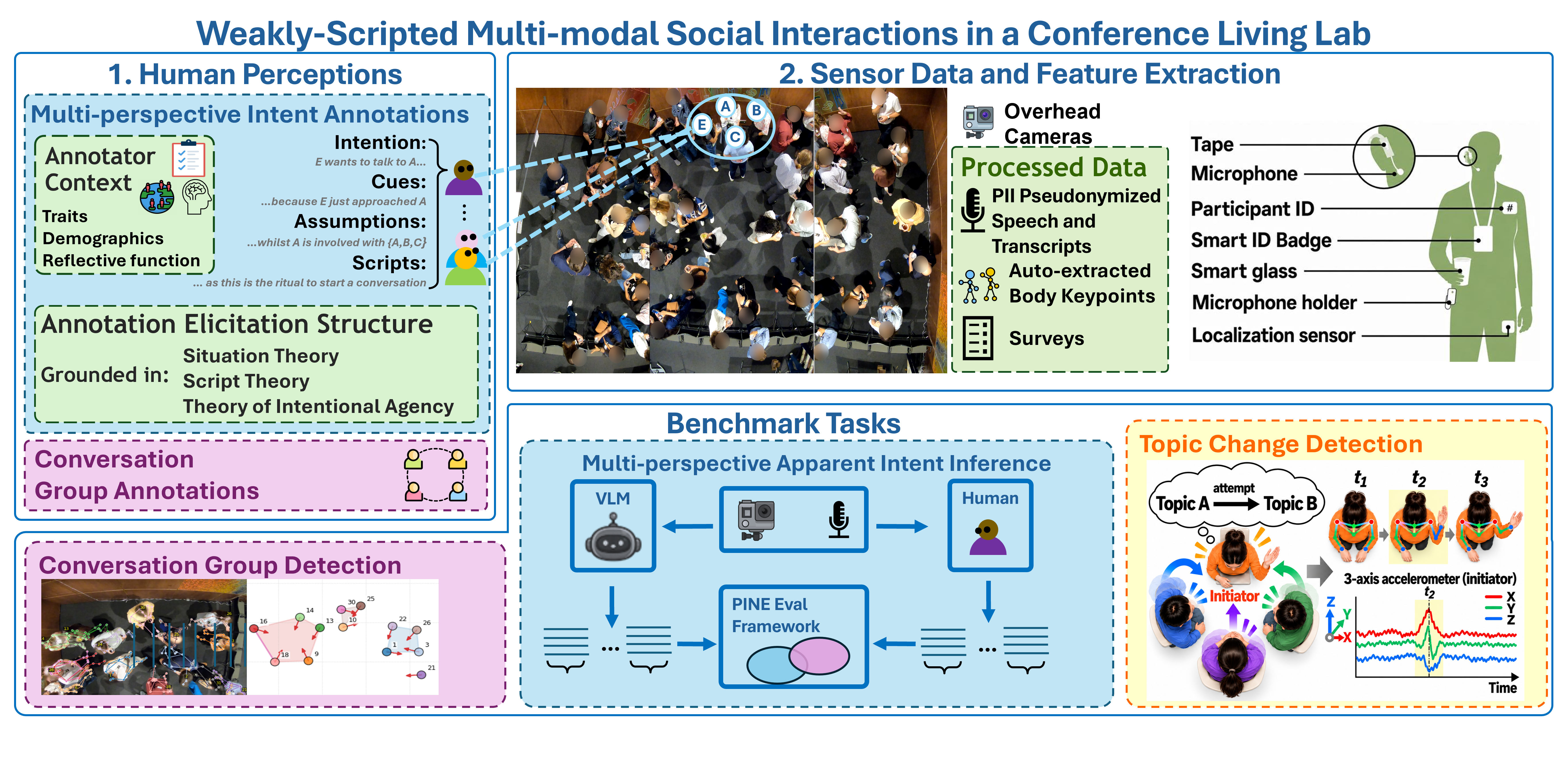}
    \caption{COSI-Lab: The conference living lab for studying multi-perspective social intentions.}
    \vspace{-2pt}
    \label{fig:ingroup_teaser}
\end{figure*}


Thus far, weakly scripted multi-party datasets have only been developed for detection and localization tasks \cite{raman2022conflab,MnM2021_underline,SALSAData2015}, without access to full verbal and nonverbal information that could disentangle nuances in conversational intentions. To this end, we introduce the Conference Social Intentions Living Lab (COSI-Lab) which captures real-life weakly scripted and fully synchronized multisensor multimodal data of 32 academic participants interacting over the course of 3 hours. We record both self-reported distal intentions ($\sim 30$ minutes) (e.g. to lay foundations for future collaborations) and crowd-source third-party proximal intention perceptions ($\sim 1-30$ seconds) (e.g. to show interest in the current conversation). Aside from designing a workshop living lab, we contribute annotations and data processing for three key benchmark tasks: (i) subjective multimodal AII inference using our proposed perspectivist intention narrative evaluation (PINE) (ii) multimodal social involvement estimation, and (iii) multimodal topic change estimation.
Consequently, COSI-Lab is designed around these desiderata for (social) intent-aware systems and dataset design:


\emph{\bf{D1.} Multiple plausible intent inferences can exist simultaneously:} outputting a single true (social) intent may not always be desirable; by design, it is safer and more ethical to allow for multiple plausible hypotheses about a person's intention to be generated and used at inference time. For example, a system should reason across multiple plausible hypotheses of a human's intentions even if that human is unwilling or unable to provide their true intent to the system for any reason (e.g. practical, cognitive, or privacy constraints). 

\emph{\bf{D2.} Account for unrealized intentions not manifested in future outcomes}: Exclusively looking to the future to label intentions in the past risks biasing perceptions towards outcomes. Intentions that are never realized could affect a person in ways that would be inaccessible to the system. 

\emph{\bf{D3.} Need explanations that account for ambiguity, subjectivity, and context dependence}: how does the observable evidence available, any intermediate interpretations by a perceiver, and their own predispositions (e.g. traits and demographics) relate to their inference? We need measures to capture qualities of explanations grounded in existing theory in cognition and situational awareness.

\emph{\bf{D4. Account for the hierarchical structure of human intention:}} Individuals rarely pursue goals in isolation. Instead, their behavior is continuously shaped by nested layers of motivations at different levels of abstraction \citep{KRUGLANSKI201569}, often involving concrete proximal outcomes within a situation (e.g., "Introduce yourself to  X") in service of distal ones (e.g., "Be a successful networker"). 


\section{Related Work}
\vspace{-4pt}

Table \ref{tab:dataset-comparison} summarizes the comparison between COSI-Lab and other multimodal human behavior and intention-related datasets. 

\begin{table*}[t!]
\begin{threeparttable}
\caption{Comparison of human behavior datasets with intention and reasoning annotations.}
\label{tab:dataset-comparison}

\scriptsize
\setlength{\tabcolsep}{3pt}
\renewcommand{\arraystretch}{0.95}

\newcolumntype{Y}{>{\raggedright\arraybackslash}X}

\begin{tabularx}{\textwidth}{@{} 
>{\hsize=0.75\hsize}Y  
>{\hsize=1.05\hsize}Y  
>{\hsize=1.25\hsize}Y  
>{\hsize=1.15\hsize}Y  
>{\hsize=0.70\hsize}Y  
>{\hsize=1.10\hsize}Y  
c c c c                
@{}}

\toprule
& \multicolumn{5}{c}{\textbf{Data capture}} & \multicolumn{4}{c}{\textbf{Desiderata}} \\
\cmidrule(lr){2-6} \cmidrule(l){7-10}

\textbf{Dataset} & \textbf{Scenario (Scripting)} & \textbf{Intention Representation} & \textbf{Narrative Annotation} & \textbf{Modalities} & \textbf{Annotation Size} & \textbf{D1} & \textbf{D2} & \textbf{D3} & \textbf{D4} \\

\midrule

PIE++ \citep{khindkar2024can} & Pedestrian-Vehicle (Strong) & Categorical & Multi-annot. prediction & Video & 5k instances of 1-2 sec. &\cmark &\cmark &\xmark & N/A \\
\addlinespace[1.5pt]

PSI \citep{jingpsi} & Pedestrian-Vehicle (Strong) & Categorical Free-form & Multi-annot. subjective & Video & 196 vids of 15 sec. &\cmark &\cmark & \cmark & N/A \\
\addlinespace[1.5pt]

IntentQA \citep{li2023intentqa} & Skilled activities (Strong) & Categorical guided Q\&A & Multi-annot. agreement & Video & 5k vids of 44 sec. &\xmark &(\cmark) &(\cmark) &\cmark \\
\addlinespace[1.5pt]

Ego-Exo4D \citep{grauman2024ego} & Skilled activities (Strong) & Expert free-form/reflections & Multi-perspective \& Multi-annotator & Video, eye, IMU & 5k takes of 2.6 min. &\cmark &\xmark & \cmark& \cmark\\
\addlinespace[1.5pt]

I2M \citep{umagami2026intend} & Daily living (Strong) & Free-form descriptions & Single-perspective logs & Video, IMU & 215 seqs of 2.8 min. &\xmark &\xmark & \xmark& \cmark\\
\addlinespace[1.5pt]

MIntRec2.0 \citep{zhang2024mintrec} & TV shows (Full) & Categorical, no reasoning & Multi-annotator agreement & Video & 1.2k dialogues of 3 sec. & \xmark& \xmark&\xmark &\xmark \\
\addlinespace[1.5pt]

\textbf{Ours (COSI-Lab)} & Social interaction (Weak) & Structured Free-form \& counterfactual & Subjective/Multi-annot. & Video, Audio, IMU & 570 vids of 30s (4.6k annotations) & \cmark& \cmark&\cmark &\cmark \\

\bottomrule
\end{tabularx}
\begin{tablenotes}[flushleft]
    \tiny 
    \setlength{\itemindent}{-3pt}
\item(\cmark): no explicit stance is given in paper; we provide a best guess based on partial or implicit fulfillment. N/A: it was not relevant to intended scope of paper. 
\end{tablenotes}
\end{threeparttable}
\end{table*}

\emph{\bf{Outcome-oriented pedestrian intention.}} Pedestrian intention datasets such as PIE \citep{rasouli2019pie} formulate intention as short-term outcomes. PSI \citep{jingpsi} enriches intention labels with multiple subjective, free form explanations from the driver perspective. However, most other datasets rely on strongly scripted settings where intentions are prescribed with a few classes \citep{khindkar2024can, jingpsi}. 

\emph{\bf{Reasoning of intention.}}   IntentQA \citep{li2023intentqa} uses a guided question-answering approach to probe the intention behind daily actions. However, the multiple-choice format limits intention labels to a single perspective. Ego-Exo4D \citep{grauman2024ego} provides expert descriptions with first-person reflections on skilled activities. Recent work has applied MLLMs to evaluate social understanding \citep{zhou_sotopia_2024}, multiparty conversation \citep{wu_how_2025}, affect perception \citep{lian_affectgpt_2025}, multimodal intention recognition \citep{zhang_can_2025, peng2026eye}, and reasoning traces during social inference \citep{mathur_social_2025}. However, these datasets and benchmarks predominantly focus on predefined tasks where the structure of activity is strongly-scripted in nature, limiting the possible emergent intentions that could arise more open weakly scripted settings. 

\emph{\bf{Multimodal human behavior.}} Multimodal behavior datasets \citep{raman2022conflab, cabrera2018matchnmingle, agrawal2025seamless} are typically not focused on social intentions, and also lack either ecological validity or broad coverage of modalities. MIntRec2.0 \citep{zhang2024mintrec} labels categorical social intentions in scripted TV dialogue, limiting analysis of spontaneous interaction. PARSEL \citep{hrkalovic2025parsel} captures audiovisual dyadic conversations and cooperative partner choices, but represents a narrower (pre-arranged dyadic-only) setting for partner selection than COSI-Lab.

\vspace{-2pt}
\section{Data Collection and Processing}
\vspace{-2pt}

\vspace{-2pt}
\subsection{Procedure Design} 
\vspace{-2pt}

We developed a multi-stage data collection framework that captures the interplay between stable individual traits, dynamically evolving individual goals and affective states, and multimodal networking behaviors. Data were gathered as part of a pre-conference workshop  to maximize the ecological validity of the participants' social intentions within a real networking event. Data collection venue and participant recruitment logistics was supported by the conference organization, however the board and all participants were unaware of our specific research questions.

\vspace{-2pt}
\paragraph{Participants and recruitment} We recruited 32 participants, primarily academic researchers from social or computer science, through official conference organization announcements, social media posts, and word-of-mouth. Participants were informed in advance via email and the workshop website about the study procedures, the full set of recorded modalities, and the data sharing conditions. They provided explicit consent for both recording and data sharing.

\begin{figure*}[t]
    \centering
    \includegraphics[width=\textwidth]{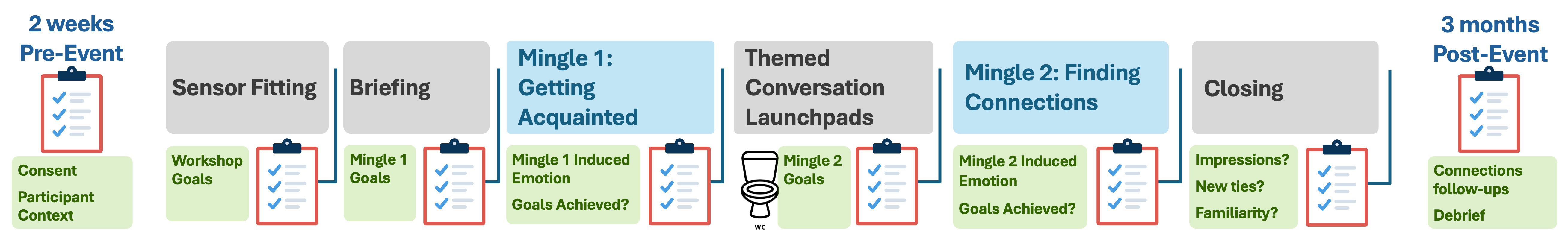}
    \caption{Workshop procedure}
    \label{fig:ingroup_workflow}
\end{figure*}

\vspace{-2pt}
\paragraph{Workshop procedure} 
The data collection followed a structured multi-stage design (see Figure~\ref{fig:ingroup_workflow}). 
During the pre-event phase, participants completed questionnaires assessing demographic information and psychometric measures to establish baseline traits that influence social intent. The in-event phase consisted of two 30-minute mingling sessions where goals, reflections on goals, and affective states \citep{broekens2013affectbutton} were provided by participants before and after each mingle. A post in-event survey captured interaction impressions, intent to connect, and interpersonal acquaintanceship. three months later, a post-event survey was used to identify realised new connections.

\vspace{-2pt}
\paragraph{Trait and Demographics.}
Surveys were conducted at different stages of the workshop (see  figure~\ref{fig:ingroup_workflow}). Specifically, self-monitoring personality \citep{lennox1984revision}, social anxiety \citep{connor2000psychometric}, communication and interaction styles \citep{vissa2012agency, rubin1994development, cheek1981shyness}, and HEXACO personality traits \citep{ashton2009hexaco}). Items were aggregated into one value per measure per participant. 
To reduce identifiability, age was quantized, and country of birth, country of residence, nationality, and ethnicity were collapsed into broader categories where appropriate.  

\vspace{-2pt}
\subsection{Data Capture Setup and Post Processing} \label{sec:data_capture}
\vspace{-2pt}


We record multimodal synchronized data for the participants during the 2 mingle sessions. We extended the privacy-by-design fully-synchronized capture set-up of \citet{raman2022conflab,Raman2020} by adding a smart cup, speech quality microphones and ultra-wideband (UWB) wearable badges as described below. 
See synchronization method and details of the data capture in Appendix Section~\ref{sec:app-data-collection-setup}.


\vspace{-2pt}
\paragraph{Cameras.} Video was recorded with GoPro Hero 7 cameras. In both sessions, 5 ceiling mounted cameras provided overhead views and 4 corner cameras provided elevated side views. Due to privacy considerations related to facial visibility, only overhead views are included in the dataset. Cameras are calibrated by estimating intrinsics from ChArUco-board videos and, after installation, extrinsics from synchronized footage of ChArUco boards moved through the shared interaction space. 
In post processing, individuals were segmented in each frame using SAM3~\citep{carion2025sam}, and the resulting masks guided ViTPose~\citep{xu2022vitpose} to detect 2D body keypoints. Using the estimated camera parameters, we reconstructed these keypoints in the shared 3D room coordinate frame, which gives 3D keypoint extraction. Further details are provided in Sections~\ref{subsec:calibration_details} and~\ref{subsec:3d_kp_extraction}, with example calibration footage shown in Figure~\ref{fig:calibration}.



\vspace{-2pt}
\paragraph{Smart Wearables.} The Midge ~\citep{Midge} is an open-source wireless wearable with a microphone, 9-axis Inertial Measurement Unit, and Bluetooth for proximity sensing. Each participant wore one hung around their neck like a conference ID badge. A second one was attached to a drinking glass, which they could use during the workshop. To obtain an absolute position in 3D space, the participants were also fitted with UWB badges placed in a pocket. We used 32 Sewio Leonardo Personal tags~\cite{LocatifyTag} and 5 Vista Omni anchors~\cite{LocatifyAnchor}. Note that UWB indoor positioning does not work well in high density situations, despite commercial claims (see preliminary error estimation in Appendix~\ref{app:processing}). Audio recorded by the smart wearables were pseudonymized (see description below).



\vspace{-2pt}
\paragraph{Microphones.}High-quality audio was captured using 32 Sennheiser wireless microphones attached to a participants' cheek with Lavalier tape. For audio pseudonymization, full recordings were transcribed directly without speaker diarization to preserve original audio. PII detection was performed using Microsoft Presidio~\citep{asimopoulos2024benchmarking}, which has performed well in prior benchmarks. Detected names, email addresses, and phone numbers were mapped back to their corresponding audio intervals and rendered unintelligible using a 1.25~kHz low-pass filter. For the mingle sessions, speaker diarization was additionally performed to identify and isolate the microphone wearer, from which participant-specific pseudonymized transcripts were generated.

 \vspace{-2pt}
\section{Data Annotation Design} \label{subsec:data_annotation}
\vspace{-2pt}
\subsection{Design of apparent intent annotation process}
\vspace{-2pt}

As mentioned before, AII is inherently subjective; different observers may construct different interpretations of the same situation based on personal differences, related to experiences, abilities, cultural background, or personal disposition.
Moreover, participants may maintain multiple concurrent proximal intentions, each influencing their planning behavior as they work toward particular goals and sub-goals\citep{KRUGLANSKI201569}. This leads to open challenges for evaluating the multiplicity of plausible intention narratives that could be perceived by external observers.


\paragraph{Reasoning about intentions in weakly scripted settings} To reason about intent narratives, we leverage situation \citep{RAUTHMANN2021427} and script  \citep{scripttheory} theory. These are the observable \emph{cues} of the scene and assumptions individuals make about how a situation unfolds. Using the 3C's taxonomy \citep{RAUTHMANN2021427}: \emph{Cues} are directly perceivable elements such as actions, gestures, speech, objects, sounds, and spatial context; \emph{characteristics}, which reflect the perceived affective qualities of the interaction inferred from cues (e.g., tense or friendly); and \emph{classes} refer to the situation type including expectations about roles and relationships (e.g., a job interview, casual conversation, or negotiation). Finally, \emph{social scripts} \citep{scripttheory} describe the expected sequence of actions performed by people in the situation as well as the role they play. This implicit “playbook” guides how interactions typically proceed, such as greeting rituals, introducing, conversing, and closing. Together, cues and these layered assumptions shape the intention narratives. On top of this, the annotators were instructed to consider cues, situation characteristics, situation class, and social scripts.
\paragraph{Intention annotation procedure}
To address the nature of subjectivity in narrative construction, Analysis-Holism Scale\citep{MARTINFERNANDEZ2022111322} and Reflective Functioning Questionnaire \citep{fonagy2016development} questionnaires were completed in addition to surveys of same measures as the participant pre-workshop survey. This allows for annotator and participant profile comparison analysis. The annotation procedure was as follows. Six annotators observed a $30$-second overhead video clip of a conversation with the speech for all conversation participants. Each annotator was assigned a specific participant and asked to describe the intention they perceive that individual to have at a given moment. They were then asked to justify their interpretation, provide confidence scores for both the intention and its explanation, and rate the intensity of the intention (i.e., how strongly it appears to drive the participant’s behavior). They marked in time the start and end of the intention. Multiple concurrent intent annotations as well as changing intentions could also be annotated. 

\paragraph{Pilot and Full Study} A pilot study (n=48) explored how two varying levels of instruction affected annotations; fully free-form (A) vs. structured free-form (B). While (A) led to confusion, both conditions revealed that annotators did not naturally incorporate explicit assumptions into their explanations. Thus, final instructions were updated to explicitly define the relevant cues and assumptions, requiring annotators to incorporate them into their reasoning. The final instruction set and both pilot instructions can be found in Appendix \ref{app:annotation} with a description of the annotation process. The full study contained 65 annotators and 7,176 human annotations.
Annotators were recruited via Prolific and completed the task online using a custom annotation interface adapted from Covfee \citep{CovfeeGithub}.


\subsection{Design of conversation group annotation} \label{subsec:conversation_group_annotation}


Social conversational involvement is a key contextual scoping factor demarcating where social influence and intent might exist. To contextualize social intention within an interaction, we consider the notion of conversation floors, which could be approximated by a group of people sharing a common spatial attentional focus~\citep{edelskyWhoGotFloor1981}. To operationalize this joint attention focus, we also take inspiration from Kendon's theory on F-formations~\citep{KendonInteraction1990} which translates this group into physical arrangements. Conversation groups are crucial to understand social involvement and social intentions, as they define the spatial and social boundaries of ``who is talking with whom''. Crucially, we depart from prior work which labeled F-formations identified as groups who stand with their lower body facing each other \citep{hung_detecting_2011,MnM2021_underline,raman2022conflab}. Instead, conversation groups were annotated by considering jointly focused head orientations. Annotations were obtained for both mingle sessions by marking any \textit{changes} in each conversation group, including when a composition change occurred as well as the new group members. This yielded \textit{continuous} annotations at the same granularity as camera frame rate (59.94Hz).

\subsection{LLM-based topic redirection labeling} 
\label{subsec:topic_change_annotation}
In natural multiparty conversations, participants may attempt to change the topic as part of their ongoing goals. Such attempts may not always be taken up by others. The ability to avoid or enforce certain topics can be a sign of wishing to avoid emotional discomfort \citep{Caughlin2004}, assert dominance \citep{ReidNg2000}, or manage self-presentation  \citep{Holtgraves1992}.  Following \citet{button1985topic,schegloff2007sequence}, we label \emph{topic-redirection attempts} separately from their subsequent \emph{uptake}. Labels were auto-generated from pseudonymized transcripts using a locally hosted Gemma 4 31B model in two stages. First, each target turn was judged using only the preceding context and the current turn; later outcomes did not influence whether a redirection attempt was identified. Then, the  identified attempt and its proposed topic were stored. The second stage examined the subsequent conversation to determine whether the group took up that topic. The resulting labels are used in both Fig.~\ref{fig:topic-uptake-by-group-size}B and Benchmark~3 (Section~\ref{subsec:topic_change_detection}); full procedure and prompts are provided in Appendix~\ref{sub:topic_redirection_coding}.

\vspace{-2pt}
\section{Descriptive Statistics}
\label{sec:descriptive_statistics}
\vspace{-2pt}

\begin{figure*}[t]
    \centering
    \includegraphics[width=\textwidth]{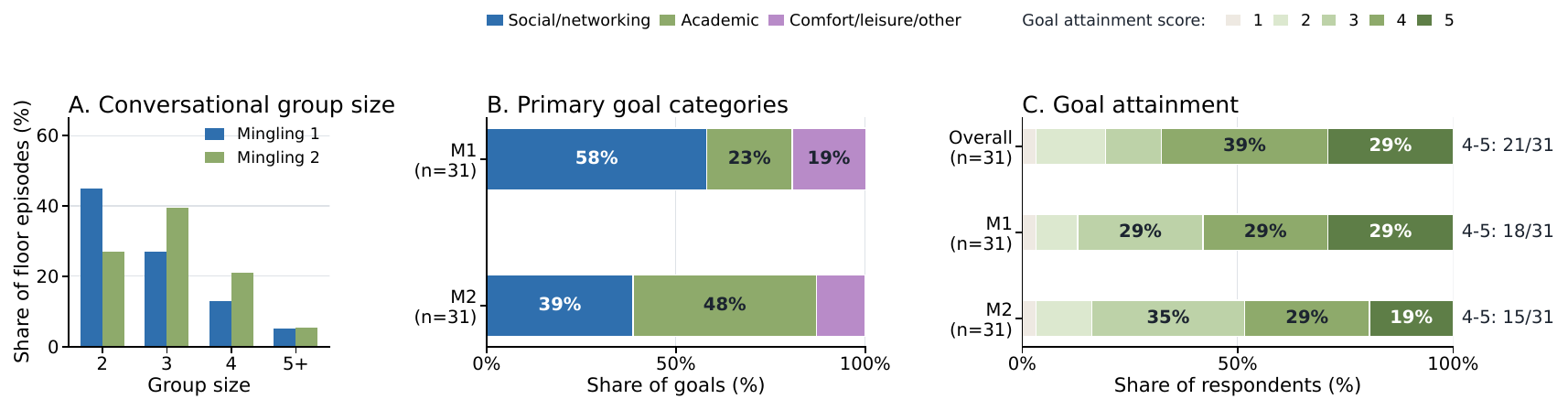}
    \caption{Interaction structure and survey-reported goals across the two mingling sessions.}
    \label{fig:descriptive-interaction-survey-overview}
\end{figure*}

\begin{figure*}[t]
    \centering
    \includegraphics[width=\textwidth]{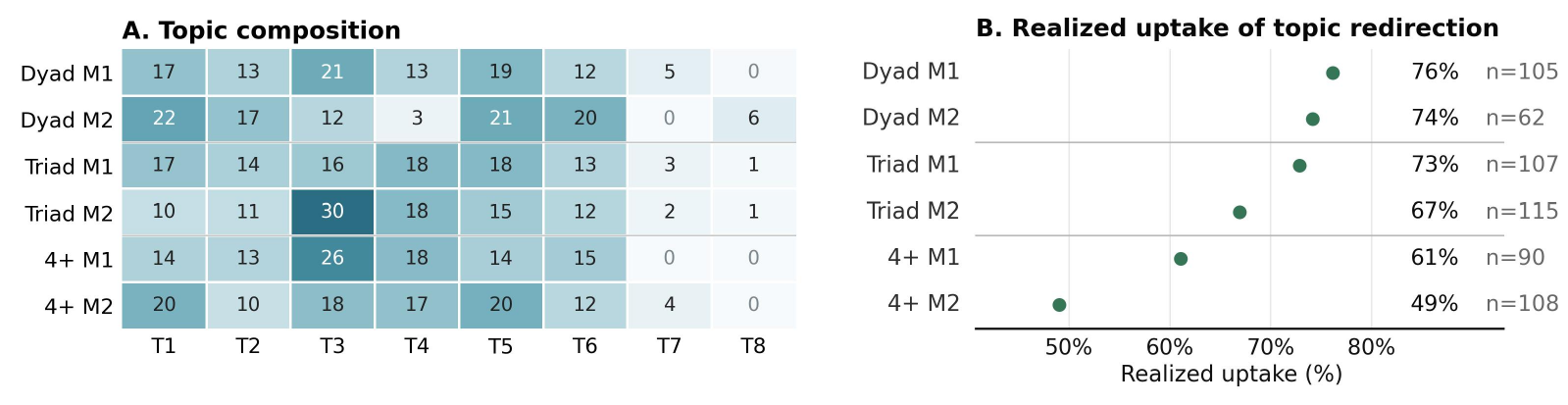}
    \caption{Conversation topic composition and redirection uptake by session and  group size. M1/M2 denotes Mingle~1/2. Panel~A: topic-family shares (T1: background; T2: research domains; T3: methods/data; T4: team/organization; T5: networking; T6: personal; T7: gender/diversity; T8: other). Panel~B: realized uptake among valid topic-redirection attempts. \vspace{-4pt}}
    \label{fig:topic-uptake-by-group-size}
\end{figure*}

\paragraph{Annotator profile varies intent reasoning}
To test whether who an annotator is shapes the intention narratives they
produce, we linked surveys to annotations for 52 annotators (4{,}262 narratives
over 568 clips). Each pair of annotators sharing a clip contributes the semantic similarity of their narratives, regressed on the Euclidean distance between their standardised 29-dimensional survey vectors, in a mixed model with clip and annotator as crossed random effects, the annotator entering for both members of each pair \citep{baayen2008crossed,browne2001mmmc}, fitted by REML in \texttt{lme4} \citep{bates2015lme4}. The model was then compared against 2{,}000 permutations of whole survey profiles across the 52 annotators \citep{winkler2014permutation}. Shuffling at the annotator level keeps the clip and co-annotation structure fixed as well as the correlation within the survey scales, while breaking the profile--narrative link. We found that annotators with more similar profiles wrote more similar \emph{explanations} of intention. Moving one standard deviation (SD) from the mean profile lowers narrative similarity by $\beta = 0.0102$ (SE $= 0.0025$), a fall of $0.23$ residual SD across the 10th-to-90th-percentile range of profile distance. This holds under all three narrative similarity metrics (permutation $p = .009$, $.007$ and $.011$ for cosine, overlap \citep{parupudi2025magnitudematterssuperiorclass} and HTS). Annotator identity owns $29\%$ of the variance in explanation similarity against $20\%$ for the detected intention.

\paragraph{Conversational group structure and survey goals}
Figure~\ref{fig:descriptive-interaction-survey-overview} compares annotated conversation-group structure with survey-based primary goal categories and self-rated goal attainment. Mingle 1 was more dyadic, whereas Mingle 2 contained more three-person and larger conversational groups. Among 31 respondents, primary goals differed descriptively across phases, from mostly social/networking in Mingle 1 to more academic or targeted in Mingle 2. On the 1--5 goal-attainment scale, mean scores were $3.71 (1.10)$ for Mingle 1 and $3.48 (1.06)$ for Mingle 2; ratings of $4--5$ were given by $18/31$ and $15/31$, respectively. Linking these stated goals to the annotated intentions, only $76.7$ \% of annotated participant intentions matched at least one of their reported goals, with a higher match rate for participants who reported partially or fully achieving their goals ($77.6$\%) verses those who reported not achieving them ($73.2$\%). This could be an effect of the specificity of self-reports.

\paragraph{Conversational topics and redirection uptake}
Figure~\ref{fig:topic-uptake-by-group-size} provides an  LLM-assisted summary of topic composition and realized uptake of topic-redirection attempts by session and group size. Topic composition varied across both sessions and group sizes, with uptake generally higher in smaller groups. The LLM-assisted annotations were evaluated against adjudicated human annotations on stratified subsets. Topic-composition estimates had a mean absolute error of 12.1 percentage points, while the redirection labels achieved accuracy/macro-F1 scores of 0.800/0.762 for attempt identification and 0.909/0.906 for uptake outcome classification. Details of the human-validation procedure and results are provided in Appendix~\ref{appsubsec:human_validation}.

\paragraph{Reported ties over time}
Surveys captured self-reported ties before, during, and after the workshop. Respondents reported knowing 3.8 (out of 32) attendees on average before the event, mostly as weak ties. During the workshop, they reported 14.9 interaction partners on average, 78.0\% of whom were previously unfamiliar. This pattern is notable because people in networking settings often remain with people they already know~\citep{ingram2007people}. 
Three months later, post-event respondents recalled on average 9.5 other attendees and maintaining contact with 2.4 people.

\vspace{-2pt}
\section{Benchmark Tasks}
\label{sec:benchmarktasks}
\vspace{-2pt}
\subsection{Intention estimation} \label{subsec:intention_estimation_b1}
\vspace{-2pt}
\begin{figure*}[t]
    \centering
    \includegraphics[width=\textwidth]{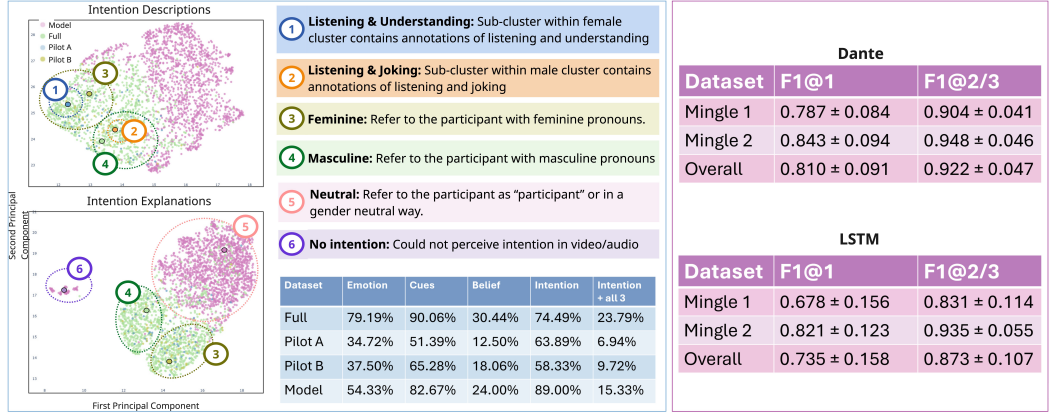}
    \caption{Left: Semantic similarity of the annotations and presence of emotions, cues, beliefs, intentions, and all 4 simultaneously.
    Right: Conversation group detection results (mean $\pm$ std over camera-wise 5-fold tests) and successful
topic-transition detection (mean $\pm$ std over five folds).}
    \label{fig:annotation-results}
\end{figure*}

\paragraph{Qualitative analysis of human and model narratives} 

Intention narratives can be analyzed qualitatively:- 
First, to assess the diverse representation of human-perceived intentions, we measure semantic similarity across annotations using Hyperbolic Tangent Similarity (HTS) \citep{parupudi2025magnitudematterssuperiorclass}. Second, we analyze narrative specificity by automatically parsing the narratives to identify the presence of dependent variables aligned with cues, assumptions, and beliefs. Third, for a subset of intention annotations, we assess narrative grounding by manually verifying whether the cues referenced in each annotation are present in the corresponding video clip or are hallucinated.

We zero-shot prompted VLMs (Gemma3 E4B \citep{gemmateam2025gemma3technicalreport} and Qwen 3 Omni 30B Instruct \citep{xu2025qwen3omnitechnicalreport}) with the personas of the annotators, built from their survey responses, and the same annotation task, asking it to generate the intention narrative that an annotator with this persona would produce. Figure \ref{fig:annotation-results}(Left) presents the semantic similarity of intention detections(top) and their explanations (bottom), comparing model outputs (gemma3: blue, qwen3: pink) with human annotations (green). A clear clustering separation emerges between model and human outputs, with overlaps especially in intention explanations, while all human annotations, occupy a broadly shared semantic space. Across both plots, annotations further group into three main clusters based on reference style: male pronouns, female pronouns, and gender-neutral terms. Human annotators predominantly use gendered references, whereas the model more often defaults to neutral phrasing, marking a key distinction. Smaller sub-clusters can also be discerned revealing similarities in perceived cues, intentions and intention reasoning.


\paragraph{Specificity analysis of human and model narratives}
The table in Figure \ref{fig:annotation-results}(Left) shows how often key components: cues, beliefs, emotions, and goal-directed intentions appear in annotations. Each percentage indicates the proportion of annotations that mention the component at least once (parsed by Gemma 4 31B \citep{gemmateam2026gemma4technicalreport}). Setting intentions aside, the model distribution of components broadly followed that of the human annotators, all three referring more often to cues and perceived emotions than to beliefs, though the models attributed beliefs more frequently and grounded their narratives in observable cues somewhat less often. Human annotators specifically focused on cues, having intentions mostly refer to actions (both video and audio) present in the clip. The models were far more explicit about intention, stating a goal in almost every narrative. This contrast arises because the models followed the instructions more closely and the implied intentions in the human annotations were sometimes missed by the LLM parser. Together, these counts give a first, coarse sense of the coverage gap between model and human narratives that we quantify more directly below.

\paragraph{Evaluating model generations} Our Perspectivist Intention Narrative Evaluation (PINE) framework compares model and human narratives along three dimensions: \emph{grounding}, \emph{human alignment}, and \emph{coverage}. First, applying the grounding test to a sample of intention narrative annotations ($n=83$), every human narrative referenced cues genuinely present in the clip, as did $97.6\%$ of model narratives. The models nonetheless frequently misattributed a cue to the wrong participant and relied almost exclusively on audio, seemingly unable to make much use of the overhead video.

Second, we assess \emph{human alignment}. Since multiple perspectives on the same situation can be equally valid, there is no single ground-truth narrative to score against. We instead ask whether each persona-conditioned model generation is close to at least one plausible human perspective, scoring it against every annotation that human annotators produced for the same video. Our \emph{Persona Alignment Score (PAS)} thus evaluates model performance by embedding all narratives and measuring the semantic similarity between model and human embeddings using HTS \citep{parupudi2025magnitudematterssuperiorclass}. For each model narrative we keep the similarity of its nearest human neighbor, averaged over all $N$ generations: $\text{PAS} = \frac{1}{N}\sum_{i=1}^{N} \max_j \text{HTS}(v_i, h_{ij})$, where $v_i$ is the $i^{th}$ model narrative and $h_{ij}$ the $j^{th}$ human annotation of the same video  (validation of score in Appendix~\ref{app:pas-alginment}). Running this comparison across both models, we found that qwen3 scored higher than gemma3 on both narrative fields (PAS $0.419$ vs.\ $0.388$ for intention detections and $0.438$ vs.\ $0.416$ for explanations), though both remained below the agreement reached between independent human annotators of the same video ($0.492$ and $0.494$ respectively).

Third, \emph{coverage} assesses whether the model's generations span the human distribution of perspectives of the given dataset without collapsing onto a narrow subset or spilling outside it. Thresholding pairwise HTS similarities, calibrated on human-human nearest-neighbor distances ($0.499$ for intention detections, $0.509$ for explanations), we compute precision (proportion of VLM outputs matched to a human annotation, guarding against generations outside the human distribution) and recall (proportion of human perspectives matched to a VLM output, guarding against collapse). We complement this pairwise view with a distributional comparison (Wasserstein distance and MMD between human and model embeddings). Both models cleared this bar rarely: precision reached $0.175$ (gemma3) and $0.232$ (qwen3) for detections and $0.197$ / $0.258$ for explanations, with recall lower still at $0.117$ / $0.198$ and $0.133$ / $0.224$, qwen3 being the stronger of the two throughout and most markedly in recall. The distributional comparison separated the models far less (Wasserstein $0.019$ vs.\ $0.018$ for detections and $0.020$ vs.\ $0.016$ for explanations; MMD $0.103$ vs.\ $0.093$ and $0.089$ vs.\ $0.061$), 
indicating that model and human embeddings span comparable overall distributions even where they separate locally (Figure~\ref{fig:annotation-results}), individual generations seldom matching any single human perspective at the level humans reach with each other.


\vspace{-2pt}
\subsection{Conversation group detection} 
\vspace{-2pt}\label{subsec:conversation_group_detection}



As described in section~\ref{subsec:data_annotation}, we run conversation group detection tasks to scope the context for social intentions. Following previous works~\citep{hung_detecting_2011, Setti2015FFormationDI, vascon_detecting_2016}, we formulate group detection by leveraging each person's features, and most typically, their positions and orientations. We evaluate two baseline architectures: the graph-based DANTE \citep{swofford2020improving} and a temporal LSTM-based model \citep{tan2022conversation}, by employing the commonly used F1 metric~\citep{vascon_detecting_2016}, and measure it under threshold $T=1$. We construct the features (positions and orientations) from raw videos with SAM3~\citep{carion2025sam} and ViTPose~\citep{xu2022vitpose} models, and provide the details of this process in Appendix~\ref{subsec:3d_kp_extraction} and~\ref{appsubsec:benchmark_2}. The detection results are compared to the conversation group annotations in Section~\ref{subsec:conversation_group_annotation}. Figure ~\ref{fig:annotation-results}(Right) shows the results; the performances are high on F1. While the high scores on the baselines don't necessarily imply that the keypoints are accurate in representing the spatial coordinates, it does demonstrate that extracted spatial relationship across cameras and events are consistent and transferrable. We report more results including a cross-camera train-test setup in Appendix~\ref{appsubsec:benchmark_2}, with stable results as well. 

\vspace{-2pt}
\subsection{Topic change detection} 
\vspace{-2pt}\label{subsec:topic_change_detection}
Prior work has shown that gesture, gaze, and posture can reflect discourse structure and help identify discourse shifts \citep{quek2000gesture,cassell2001non}. Motivated by this, we examine whether successful topic transitions can be recognized from privacy-preserving body keypoints and wearable acceleration. For this task, we formulate a binary classification task to distinguish successful topic transitions from non-transitions, where
the negative class includes both unrealized redirection attempts and ordinary
conversation. We compare logistic regression (LR) and a temporal convolutional network (TCN)
using the initiator's nonverbal behavior, and a role-aware TCN--DeepSets model
that additionally incorporates the behavior of the other conversation members.
Each baseline is evaluated using upper body keypoints, and/or acceleration
from the smart conference badge with five-fold evaluation.
Figure~\ref{fig:annotation-results} reports AUROC for the successful topic
transition detection task. Benchmark details, further results, ablation studies,
and a secondary attempt-detection task are provided in
Appendix~\ref{appsubsec:benchmark_3}. Overall, the results suggest that successful topic transitions are reflected, albeit modestly, in privacy-preserving nonverbal behavior.

\vspace{-2pt}
\section{Limitations}
\vspace{-2pt}

\paragraph{Non-overt intentions and temporal limits} While we account for the multiplicity of overt intentions, our framework does not yet account for intentions that may not be observable from observable planning behavior, either within or beyond the bounds of our 30s segments. While intent-outcome relationships can be captured within these shorter segments and the segmentation is designed to enable independent samples for training models, relationships spanning multiple segments may be missing.  We therefore warn potential users to be mindful of this in future work on longer term AII.

\vspace{-2pt}
\paragraph{Long term observee self-reported goals and short term third-party annotated intentions} While COSI-lab enables initial studies linking distal intentions and outcomes with proximal ones, future users must be mindful of the conceptual distinction between self-reports and third-party annotations. Neither is a definitive representation of the truth. Rather, they can be used for developing hybrid intelligent systems \citep{akata2020research} where humans and agents jointly locate the truth.

 \textbf{Deeper investigation of low (wearable) and high fidelity data} A full investigation across modalities and sensors was not investigated and is beyond the scope of this current paper. We encourage future users of COSI-Lab to exploit and investigate the benefits of this parallel dataset; enabling training with high fidelity while having access only to low fidelity at test time. 


\paragraph{Annotator Population Distribution} Our Prolific annotator pool may restrict the perspectives captured in AII. However, PINE compares models conditioned on the same personas against the human distribution, remaining valid for evaluating alignment with the represented population.

\vspace{-2pt}
\section{Conclusion} 
\vspace{-2pt}
The presented COSI-Lab dataset multisensor multimodal dataset of a real scientific workshop consisting of 32 participants of varying seniority and acquaintanceships. It enables, for the first time, the study of relationships between self-reported long term goals and how they relate to apparent proximal intentions. We gathered intent narrative annotations for AII using a perspectivist approach. We provide 3 benchmarks tasks on intention estimation, social involvement identification, and topic changes. By working as a multidisciplinary team, we have produced a rich interdisciplinary resource to study unscripted human social dynamics and their relation to social intentions. 

\newpage 
\subsection*{AI disclosure}

In this work, we used generative AI tools to implement methods, assist with translation, design or provide feedback on research methodology or experiments, support qualitative and thematic data analysis, and interpret results. We did not use generative AI tools to generate synthetic datasets, clean or reformat datasets, or propose or refine hypotheses. The development of theoretical models or conceptual frameworks, formulation of mathematical claims, provision of critical ingredients for proving mathematical claims, and assistance in writing proofs were not applicable to this work.

We reviewed all AI-assisted work. We checked LLM-generated methodological suggestions against the relevant literature to identify potential overlap with existing work. We manually reviewed LLM-assisted analyses to verify their accuracy and consistency with the underlying data and research materials. All AI-generated code was reviewed, tested, and verified for correctness by at least one author. We take responsibility for the final content of this work, including any text, claims, or artifacts produced with the aid of generative AI.

\subsection*{Ethics statement}
The data collection study received ethics approval from (institutional ethics review board, obfuscated for double-blind reasons). Participants were informed about the study procedures, recorded modalities, and data sharing conditions and have provided informed consent. Annotators were recruited through Prolific and received \pounds12 per task (each task is approximately 1 hour). 

To reduce identifiability, released video is restricted to overhead views. Detected personally identifiable information (PII) is pseudonymized (see Appendix~\ref{app:processing}) and corresponding audio is filtered using a 1.25kHz low frequency to reduce intelligibility. Survey data were aggregated in post-processing (see Appendix~\ref{sec:app-data-collection-setup}). Access to raw, processed, and sample data requires signing an End User License Agreement pending review, and is limited for research purposes only.

\subsection*{Reproducibility statement}
To support reproducibility, Appendices~\ref{sec:app-data-collection-setup} and \ref{app:processing} document the data collection procedures, sensor calibration and synchronization, survey processing, PII redaction, and keypoint extraction. Appendix~\ref{app:annotation} provides the annotation instructions, procedures, and narrative analysis settings. Appendix~\ref{app:descriptive_statistics_details} contains the model prompts used in Section~\ref{sec:descriptive_statistics}. Appendix~\ref{app:benchmarktask} specifies the model configurations, evaluation protocols, and computational resources for the three benchmark tasks in Section~\ref{sec:benchmarktasks}. We will include links to the code repository and dataset repository, together with the CROISSANT metadata upon publication. \footnote{Challenges in preserving double-blind anonymity prevent us from sharing at review stage because of a data sharing protocol that requires review.}

\bibliography{references}
\bibliographystyle{iclr2027_conference}



\newpage
\appendix
\input{COSILab_appendix}
\end{document}

%% file: COSILab_appendix.tex
\appendix

\begin{center}
\Large
\textbf{COSI-Lab} \\
\vspace{0.2em}Appendices \\
\vspace{0.3em}
\end{center}

\lstdefinestyle{promptstyle}{
    basicstyle=\ttfamily\scriptsize,
    breaklines=true,
    breakatwhitespace=false,
    columns=fullflexible,
    keepspaces=true,
    frame=single,
    framerule=0.25pt,
    rulecolor=\color{black!35},
    xleftmargin=0.5em,
    xrightmargin=0.5em,
    aboveskip=0.65em,
    belowskip=0.65em,
    captionpos=t,
    numbers=none,
    showstringspaces=false
}

The Appendices sections are inspired by the Datasheet for Datasets \citep{gebru2021datasheets} and are organized as follows:


\section{Dataset documentation} \label{appsec:dataset_documentation}

\paragraph{Motivation and Intended Use} COSI-Lab is a multimodal multisensor dataset focusing on human social intentions in networking settings. The motivation of collecting this dataset is to create a resource for interdisciplinary researchers to investigate human social intentions, following annotation designs that emphasize interpretations from multiple annotators who explain perceived intentions. The dataset contains 32 participants. The associated time-code synchronized modalities include 10 overhead cameras and 4 elevated side-view cameras, high-fidelity personalized audio at 48kHz, a combination of 9DOF inertial measurement unit data, audio data at 20kHz, and Bluetooth proximity data in both a single-worn wearable badge and a sensorized cup, and lastly a Ultra wide band localization sensor data. 

The intended use of this dataset is to support the study and modeling of human social intentions in real world, multi-party interactions. It is suited for developing and evaluating methods that integrate multimodal signals, such as vision, audio, and motion to infer underlying intentions. The multi-annotation explanations of perceived intentions provide a basis for future approaches that could account for explainable intention inference and approaches for socially assistive technologies that are capable of perspective taking. More broadly, this dataset also supports interdisciplinary research, particularly with the social science domain, to explore how intentions relate to longer term goal setting and their realization over time. 

\paragraph{Hosting, distribution, and maintenance} The dataset is hosted by 4TU.ResearchData, available at \url{https://data.4tu.nl/private_collections/X9fLjIbDFm7XrgYPbwQ-NFkPS_NSmt3lwLCfWiixNgE}.
This data hosting service provides automatic management for restricted data access, which includes version control and place to publish erratum if necessary. Other dataset-related findings will be communicated on the dataset project page (\url{https://mmmconflab.ewi.tudelft.nl/}). 
In order to download the data, the user must sign an End-User License Agreement (EULA) \url{https://data.4tu.nl/private_datasets/Ssvo69oh6N4oIWxXfRgifwuxXsWeHPMXxtAdijgfjYg} which will be reviewed by the corresponding author's team before granting access to the requesters.

\paragraph{Composition} The dataset is divided into several components:
\begin{itemize}
    \item Raw data:
    \begin{itemize}
        \item raw video, audio and wearable sensors data
    \end{itemize}
    \item Processed data:
    \begin{itemize}
        \item processed video and wearable sensors data used for annotations
        \item annotation for intentions and conversation groups
    \end{itemize}
    \item Data Samples:
    \begin{itemize}
        \item Data samples of raw video, audio, and illustration of survey questions;
        \item Demonstration of the post-processing, such as visualizations of keypoints and conversation group detection.
    \end{itemize}
    \item EULA (public):
    \begin{itemize}
        \item The End User License Agreement. One needs to sign the EULA before having access to the raw, processed, or sample data.
    \end{itemize}
\end{itemize}

\begin{figure}[t]
    \centering
    \includegraphics[width=\columnwidth]{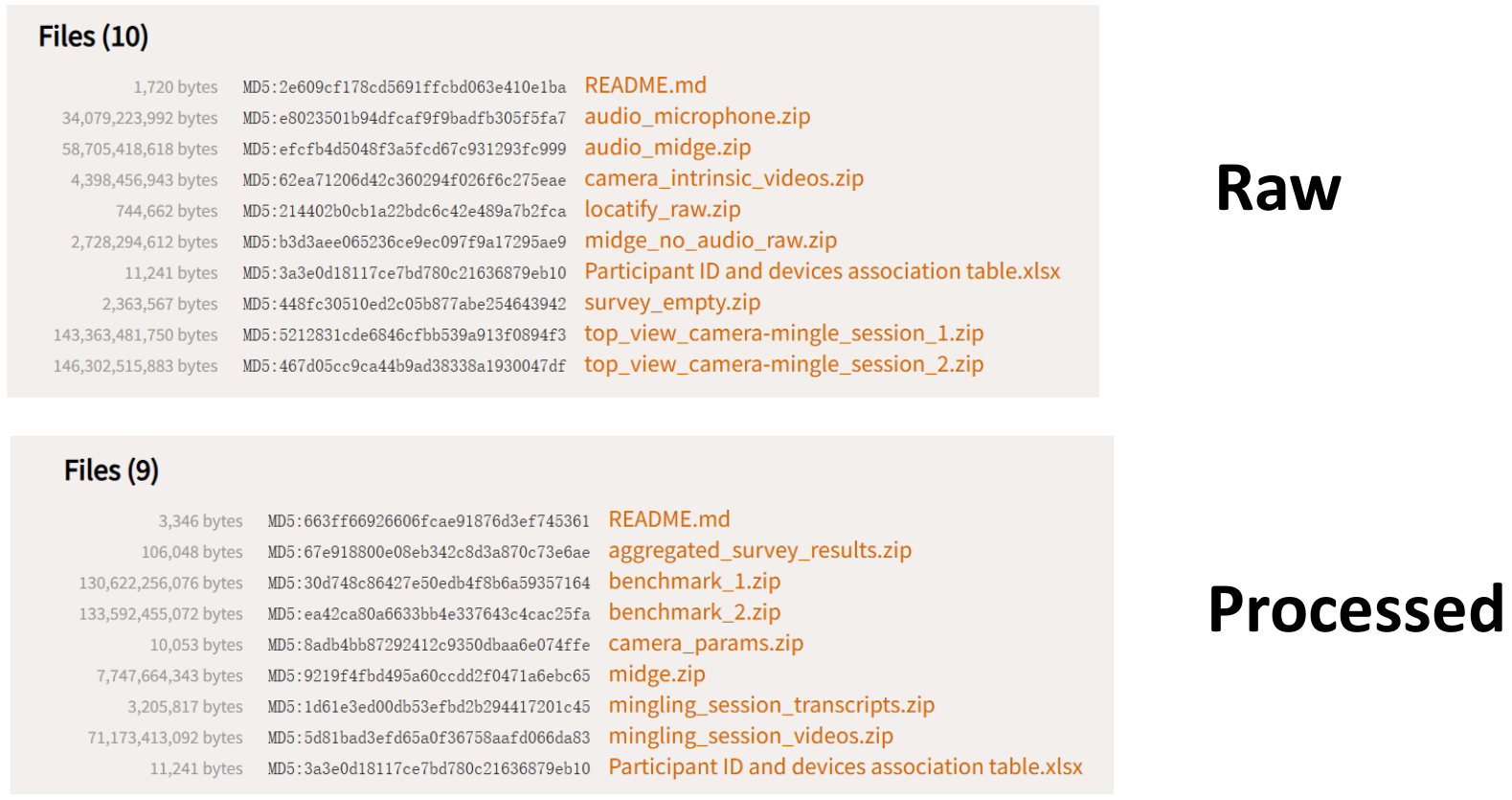}
    \caption{Dataset file structure. For more information, refer to the corresponding \texttt{README.md} files.}
    \label{fig:dataset_info}
\end{figure}

Figure\ref{fig:dataset_info} illustrates the organization of the COSI-Lab dataset. For privacy reasons, data will only be shared after signing the EULA.

\section{Details about data collection setup}
\label{sec:app-data-collection-setup}


\subsection{Details on workshop procedure design}

In the pre-event phase (two weeks prior to the event), participants completed questionnaires assessing demographic information, self-monitoring personality \citep{lennox1984revision}, social anxiety \citep{connor2000psychometric}, communication and interaction styles \citep{vissa2012agency, rubin1994development, cheek1981shyness}, and HEXACO personality traits \citep{ashton2009hexaco}. These measures provide a characterization of relatively stable individual differences that are known to influence how intentions could be formed and acted upon in social contexts. 

The in-event phase consisted of sequential components starting with arrival followed by a brief workshop introduction, then mingling session 1 (30 minutes), conversation launchpad discussion (15 minutes), mingling session 2 (30 minutes), and lastly a brief conclusion. Upon arrival, participants reported up to three workshop-level goals, ranked by importance, and indicated their affective state using the AffectButton \citep{broekens2013affectbutton}. This initial elicitation provides an anchor for individual subsequent behavior in explicitly stated goals. Participants were then equipped with sensing devices and received a briefing.

The workshop proceeded with two mingling sessions with an interleaved conversation launchpad discussion organized around thematic research topics. 
By design, the first mingling session provided opportunities to observe behaviors related to establishing acquaintances and warm-up social interactions, while the second mingling session provided opportunities for more targeted interactions that could lead to new connections.
Prior to and after each mingling session, as well as after the conclusion of the workshop, participants specified goals and reported their current affect. This setup captured the dynamic and context-dependent nature of individual goal-setting across interaction episodes, along with the perceived outcomes which could serve as a proxy signal for individual goal realization. 

In the post-event phase, participants received a follow-up email with additional details and debriefing about the sensing setup and the broader goals of the dataset. They were also asked to complete a questionnaire assessing whether they had maintained contact with, or intended to reconnect with, other participants. This measure provides an indication on long-term networking outcomes as a result of the workshop. In addition, asking participants about their level of acquaintance with others at the event allowed us to map pre-existing networks to patterns of interaction at the event.

\subsection{Sensor details}

\paragraph{Sensor settings.} The GoPro cameras were set to record at 1080p, 59.94 FPS, wide field of view, and 48 kHz audio. In the first mingle session, 5 ceiling mounted cameras provided overhead views and 4 corner cameras provided elevated side views. In the second session, 5 other ceiling cameras were used again along with the same 4 corner cameras.
The midges were set to record audio at 20 kHz. The badges were remotely started and stopped through a custom software which also controls for clock synchronization with the connected laptop. 
The Senheisser microphones were a combination of two modelds: half of which were SK 20~\citep{SennheiserSK20} and the other half were SK 2000~\citep{SennheiserSK2000}. Signals were digitized with an RME M-32 AD ~\citep{RME32AD} and recorded via an RME Fireface UFX III~\cite{RMEFireface} audio interface.

\paragraph{Time synchronization.}Time synchronization is crucial for multi-modal datasets. To align sensors, camera, and audio data, we employ the protocol with a single master source of time code. The time code is fetched from online NTP server~\footnote{https://tf.nist.gov/tf-cgi/servers.cgi} and broadcast to each sensor by :pulse. 

Midges support clock synchronization through Bluetooth connection, allowing for privacy-preserving data collection with millisecond level time sync alignment. The synchronization was done every 5 minutes. However, due to the amount of devices present, not every sync message was received every time by each midge. We observed that around a quarter of them would be reachable on each time sync round. This is still within the parameters defined in~\citep{Raman2020}.

For cameras, each camera was mounted with a Syncbac PRO to embed the time code signal into the video. For audio, the michrophones record the time code signal from the :pulse in an additional audio channel.
The 33 channels (32 microphones plus the time code) are recorded on an HDD connected to the Fireface. UWB sensors were synchronized with the other modalities by connecting the laptop running the localization software to the same network time protocol as the Midge laptop.





\paragraph{Camera calibration.}\label{subsec:calibration_details}We perform camera calibration to obtain camera intrinsic and extrinsic matrices, which are needed for downstream 3D processing tasks. We recorded a calibration section before each mingling session using several calibration boards. In previous work~\citep{raman2022conflab} the camera calibration was done by placing marks in the ground at regular intervals. This is a time consuming process, both placing the marks and manually annotating them in the recorded images. Moreover, this leads to poor camera matrix estimations, as this is equivalent to performing a camera calibration based on a single static board placed on the ground.

Initially, we attempted full bundle adjustment for the calibration. However, we found that the calibration had difficulty converging to a good optima due to the distance between the boards and the cameras. Therefore, we recorded additional videos for intrinsic camera calibration, which were used to initialize the extrinsics optimization.We employed calib.io's calibrator software~\citep{Calibrator} to run to detect the board features and run the optimization. The advantage over open source alternatives is that it supports using several calibration boards at once. We used 3 ChArUco boards. Using 3 boards decreases the total calibration recording time, compared to only using 1 board. This is required, as the cameras are relatively far away from the mingle area, which would lead to very long and tedious calibration recording sessions. 

The technical details for each board as specified in Table~\ref{tab:charuco-board-specs}, where each board has a unique starting ID for the ArUco markers: 0, 12, and 24, respectively. With this approach we obtained camera parameters that lead to an average reconstruction error of 0.8843 pixels for mingle session 1 and 0.7644 pixels for mingle session 2.

\begin{figure}[t]
    \centering
    \includegraphics[width=\columnwidth]{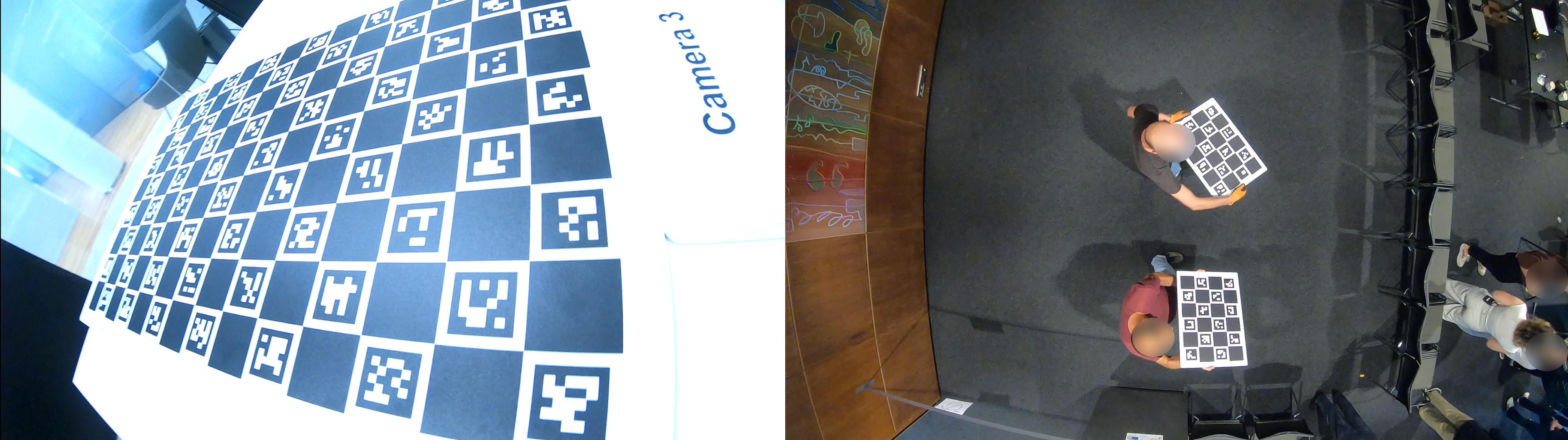}
    \caption{Example calibration footage for camera geometry estimation. The left panel shows a per-camera ChArUco-board recording for intrinsic calibration, and the right panel shows multi-camera footage used to estimate extrinsic camera poses after installation.}
    \label{fig:calibration}
\end{figure}

\subsection{Survey details}
\paragraph{Survey contents.}The questionnaires are designed based on several values. We present below the brief summary of the components in each survey. For complete set of questions, refer to the dataset link: 
\begin{itemize}
    \item \textbf{The pre-workshop survey} is given 2 weeks prior to the workshop. Content: welcome information, self-monitoring scale, social anxiety scale, networking style, shyness and sociability, communication style, HEXACO scale, and Demographic information. 
    \item \textbf{Workshop survey} is progressively collected along with the workshop. Contents: we ask for goals before each mingling session, and check whether their goals were achieved in the mingling session. Also, we check for overall goals and how they achieve them before and (immediately) after the the workshop.
    \item \textbf{Post Workshop Survey (Immediate)} is also given immediately after the workshop. We separate this from Workshop survey for privacy reasons only. It asks each participant whether, how, and why they want to keep communication with each other participant. We provide participant photos to help identification.
    \item \textbf{Post Workshop Survey (3 months)} is given 3 months later from the workshop. Contents: whether people keep communications with those they specified directly after the workshop, and for what reason.
\end{itemize}
\paragraph{Survey processing.}The raw survey responses were processed into participant-level variables before release. 
We used a manually curated codebook to map columns in the raw data to their corresponding measure, response scale, and reverse-coding indicator. 
For each multi-item measure, item-level responses were converted to numeric scores, reverse-coded when specified, and averaged across items to produce one measure-level score per participant. 
Attention checks were handled separately and summarized as both the number and proportion of correct responses.

We also applied privacy-preserving transformations to demographic fields. 
Age was quantized into bins, and demographic attributes including country of birth, country of residence, nationality, and ethnicity were recoded into broader categories to reduce identifiability. 
In addition, raw survey metadata, including timestamps, IP address, geolocation, and response identifiers, were removed from the processed dataset.

\subsection{Survey matching across waves}
\label{sec:survey_matching}

Survey responses were matched across waves using a respondent-level matching ID constructed from the combination of several self-generated ID columns. For the pre-workshop survey, this ID was constructed from \texttt{id1}, \texttt{id2}, \texttt{id3}, and \texttt{id4}. For the workshop, post-workshop, and post-event surveys, it was constructed from \texttt{birthday\_2}, \texttt{street\_2}, \texttt{em\_contact\_2}, and \texttt{place\_of\_birth\_2}.

This matching ID was used only to link survey responses from the same respondent across waves. It is separate from the A01--A32 participant labels, which were used to link participants to the event roster, microphones, and annotations.

Table~\ref{tab:survey_matching_id_availability} summarizes the availability of respondent-level matching IDs after cleaning the survey response rows. The pre-workshop survey contained 34 unique matching IDs, exceeding the 32-person event roster, while the workshop and post-event surveys contained a small number of rows with missing matching IDs.

\begin{table}[h]
\centering
\small
\caption{Availability of respondent-level matching IDs across survey waves. Cleaned response rows do not include column headers. Duplicate rows refer to extra rows sharing an already observed matching ID within the same survey.}
\label{tab:survey_matching_id_availability}
\begin{tabular}{lrrrr}
\toprule
Survey wave & Cleaned rows & Missing ID rows & Duplicate ID rows & Unique IDs \\
\midrule
Pre-workshop  & 34 & 0 & 0 & 34 \\
Workshop      & 33 & 2 & 0 & 31 \\
Post-workshop & 32 & 0 & 0 & 32 \\
Post-event    & 28 & 1 & 2 & 25 \\
\bottomrule
\end{tabular}
\end{table}

Table~\ref{tab:survey_cross_wave_matching} reports the exact cross-wave matching coverage based on this constructed ID. Fourteen respondents were matched across all four survey waves. Therefore, analyses in the main text use the largest appropriate denominator for each survey-based result, rather than restricting all results to the four-wave complete-case subset.

\begin{table}[h]
\centering
\small
\caption{Exact cross-wave matching coverage based on the constructed respondent-level matching ID.}
\label{tab:survey_cross_wave_matching}
\begin{tabular}{lr}
\toprule
Survey comparison & Matched unique IDs \\
\midrule
Pre-workshop $\cap$ workshop & 23 \\
Pre-workshop $\cap$ post-workshop & 22 \\
Pre-workshop $\cap$ post-event & 16 \\
Workshop $\cap$ post-workshop & 29 \\
Workshop $\cap$ post-event & 19 \\
Post-workshop $\cap$ post-event & 18 \\
All four surveys & 14 \\
\bottomrule
\end{tabular}
\end{table}

We used exact matching on the constructed respondent-level ID. Possible one-field mismatches were inspected but not automatically merged.

\begin{table}[]
\centering
\begin{tabular}{|l|l|}
\hline
 Width & 600 mm \\ \hline
 Height & 400 mm \\ \hline
 Rows & 4 \\ \hline
 Columns & 6  \\ \hline
 Checker width  & 90 mm \\ \hline
 Dictionary & DICT\_5X5\_100 \\ \hline
 OpenCV legacy pattern & True \\ \hline
\end{tabular}
\caption{Charuco boards details}
\label{tab:charuco-board-specs}
\end{table}

\section{Data sanitization and processing}
\label{app:processing}

\subsection{Audio data pseudonymization}
The audio privacy processing pipeline consisted of two related branches: transcript pseudonymization and audio redaction. Both branches relied on word-level transcription, which allowed detected personally identifiable information (PII) in text to be linked to the corresponding time intervals in the audio. The goal was to reduce the exposure of identifiable linguistic content while preserving the temporal structure of the recordings for interaction analysis.

For transcript pseudonymization, we used the Sennheiser microphone recordings from the mingling sessions as the input source. To obtain high-quality transcripts, the recordings were first processed with NeMo-based speaker segmentation~\citep{kuchaiev2019nemo}, after which WhisperX~\citep{bain2023whisperx} was used to generate word-level transcriptions. PII detection was then performed on the transcripts using Microsoft Presidio~\citep{microsoft_presidio} with spaCy~\citep{honnibal2020spacy} and regular-expression based recognizers. The detected PII categories included person names, email addresses, and phone numbers. In the released transcripts, person names were replaced with consistent pseudonyms, while email addresses and phone numbers were replaced with typed placeholders.

For audio redaction, the Sennheiser and midge audio recordings were processed directly without speaker separation. This choice was made because the Sennheiser recordings were already sufficiently clear, and because applying speaker separation before redaction could introduce fragmentation and unnecessary processing artifacts. Keeping the redaction pipeline closer to the original recordings also allows other researchers to apply their own post-processing methods according to their research needs. Each recording was transcribed with WhisperX, and the same PII detection procedure was applied to the resulting word-level transcript. The detected PII spans were mapped back to their corresponding time intervals in the audio, and these intervals were processed with a low-pass transformation at 1250 Hz. This transformation reduces the intelligibility of sensitive spoken content while preserving the surrounding temporal structure of the audio.

\subsection{3D keypoint extraction} \label{subsec:3d_kp_extraction}
Obtaining 3D keypoint is necessary for inferring people's positions and orientations in the space, and subsequently for downstream human conversation group analysis. In COSILab we do this in several steps: obtaining person masks, inferring 2D keypoints, transferring to 3D keypoints, and inferring positions and orientations.

\subsubsection{Obtaining person masks}
We construct a per-person per-frame object mask by the SAM3 model~\citep{carion2025sam}. SAM3 provides precise object masks with flexible inputs, such as providing one point or bounding boxes. In our case, we first manually annotate the first frames of videos with human bounding boxes, and then keep tracking and segmenting the people in the video. In practice, we clip the videos into 20-second segments, annotate for first frames of each segment, and then concatenate the masks together. This has two advantages:
\begin{itemize}
    \item It updates people composition in the scene. Since automatic detecting people would cost extra memory on GPU, we employ a tracking-only module of the SAM3 model. This means that it can only track and not detect any new person appearing in a fixed scene. By splitting a lon gvideo into short clips, we \textit{update} visible people by annotation every 20 seconds, and therefore facilitating people being tracked.
    \item It reduces memory cost significantly. SAM3 model consumes GPU memory rapidly as the length of the video grows, especially for our case where the number of people being segmented is high. In our experiment, continuously tracking 15 people in a video of over 1 minute with 1920*1080 resolution would consume more than 80G GPU memory, so cutting the video into short segments would significantly relax this constraint.
\end{itemize}

An example of one frame of segmented people in the scene is shown in figure~\ref{fig:baseline2_pipeline}(a). The segmentation is done for both mingling sessions, each session with 3 camera videos of 35 minutes. The first session of masked video is 13:45:00 to 14:20:00 (camera 6,8,10), and the second session is 14:52:00 to 15:27:00 (camera 1,3,5). This step gives us per-frame per-person masks.

The mask extraction is done on Intel Xeon Gold 6448Y 32C 2.1GHz CPU and NVIDIA Tesla A100 GPUs (80GB memory). Under this setting, it takes around 20 minutes to generate per-frame person masks for each 20 second clip, but it varies among the number of person present in the video.

\subsubsection{Inferring 2D keypoints} \label{subsec:infer_2d_kp}
\begin{figure}[t]
    \centering
    \includegraphics[width=\columnwidth]{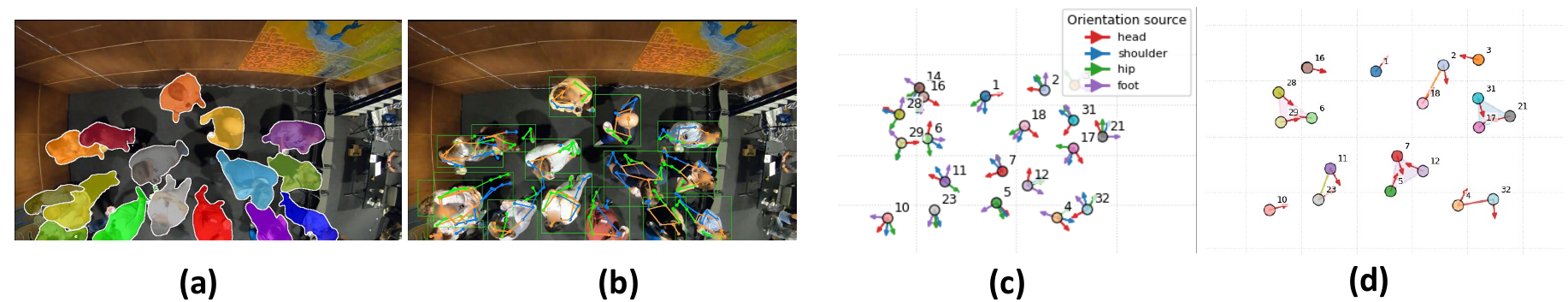}
    \caption{Illustration of the process of benchmarking task 2. (a)Segmentation mask from SAM3 model; (b)Inferred keypoints from VitPose; (c)Transferred 3D keypoints in space; (d)Illustration of the conversation group detection task.}
    \label{fig:baseline2_pipeline}
\end{figure}

Each person mask is essentially a set point. We take it's rectangular convex hull on the image as the bounding box for corresponding person, and getting per-frame per-person bounding boxes. To infer keypoints in the 2D pixel domain, we leverage ViTPose model~\citep{xu2022vitpose}. It can infer people's skeleton and provide keypoint positions in a continuous manner on a video, given the continuous bounding box information. Therefore, we use per-frame bounding boxes as input and get per-frame per-person keypoint information. An example keypoint plot is shown in figure~\ref{fig:baseline2_pipeline}(b).

The ViTPose model in this process is fine-tuned on ConfLab~\citep{raman2022conflab}, where videos from overhead cameras are accompanied with manually annotated keypoints for each person. Since the original ViTPose model was trained for detecting keypoints on the face/back side, fine-tuning on a similar setting (top-down camera) makes it perform much better on our videos.

The 2D keypoint extraction is done on AMD EPYC 7543 32-Core Processor CPU and NVIDIA A40 GPU (46GB memory). On average, one task of inferring 2D keypoints for a 5-minute video takes around 40 minutes, but also subject to number of people.

\subsubsection{Transferring to 3D keypoints}
The keypoint information obtained in~\ref{subsec:infer_2d_kp} is in 2D pixel space. However, most applications regarding conversation group analysis~\citep{swofford2020improving,tan2022conversation} require \textit{spatial} position and orientation information, i.e. within 3D world coordinates. To make the transfer, we utilize camera intrinsics and extrinsics. 

Given the camera intrinsic matrix

\[
K =
\begin{bmatrix}
f_x & 0 & c_x \\
0 & f_y & c_y \\
0 & 0 & 1
\end{bmatrix},
\]

and distortion coefficients $D$, the pixel coordinate is undistorted and converted into normalized camera coordinates:
\[
(x_i^n, y_i^n) = \mathrm{Undistort}(\tilde{p}_i, K, D).
\]

This gives a ray in the camera coordinate system:
\[
\mathbf{r}_i^c =
\begin{bmatrix}
x_i^n \\
y_i^n \\
1
\end{bmatrix}.
\]

The extrinsic parameters are represented by a rotation vector and translation vector.  
The rotation vector is converted to a rotation matrix $R$ using Rodrigues' formula.  
The extrinsic relation follows the OpenCV convention:
\[
\mathbf{X}^c = R \mathbf{X}^w + \mathbf{t},
\]
where $\mathbf{X}^w$ is a 3D point in world coordinates and $\mathbf{X}^c$ is the same point in camera coordinates.

The camera center in world coordinates is therefore $\mathbf{C}^w = -R^\top \mathbf{t}$, and the camera ray is transformed into the world coordinate system by $\mathbf{r}_i^w = R^\top \mathbf{r}_i^c.$

Each keypoint is assumed to lie at a known height above the floor.  
For body height $H$ and keypoint-specific height ratio $\alpha_i$, the world height of keypoint $i$ is $Z_i^w = H \alpha_i$. The 3D point is then obtained by intersecting the world ray with the horizontal plane $Z = Z_i^w$.  

Points on the ray are $\mathbf{X}_i^w(\lambda) = \mathbf{C}^w + \lambda \mathbf{r}_i^w.$ Solving for $\lambda$ using the $z$-coordinate gives
\[
\lambda_i = \frac{Z_i^w - C_z^w}{r_{i,z}^w}.
\]

Thus, the world position of keypoint $i$ is
\[
\mathbf{X}_i^w = \mathbf{C}^w + \frac{Z_i^w - C_z^w}{r_{i,z}^w} \mathbf{r}_i^w.
\]

and only the floor-plane coordinates are retained:
\[
(x_i^w, y_i^w).
\]

An example of the transferred keypoints in 3D is given in figure~\ref{fig:baseline2_pipeline}(c). This task does not require heavy computation resources and any commercial device would suffice.

\subsubsection{UWB to Camera Alignment and Validation}

To align UWB positions with the video recordings, we estimated a spatial transformation from the UWB coordinate system to the camera coordinate system using manually annotated point correspondences. Each correspondence paired a selected UWB location with its corresponding position in the camera image. UWB reference positions were represented in homogeneous coordinates and transformed using a 4$\times$4 matrix $\mathbf{A}$. The transformed positions were projected onto the image plane using the camera’s calibrated intrinsic parameters and fisheye distortion model, implemented through OpenCV’s fisheye projection function.

We manually annotated 16 pocket locations across participants to establish correspondences between UWB positions and image coordinates. The transformation ($\mathbf{A}$) was estimated through iterative reprojection error minimization. At each iteration, the current estimate of $\mathbf{A}$ was used to transform and project the UWB reference positions into the image plane. The reprojection error for each correspondence was defined as the Euclidean distance in pixels between the projected position and the annotated pocket location. We updated $\mathbf{A}$ using Levenberg-Marquardt optimization to minimize the sum of squared reprojection errors across all correspondences. The estimated mean physical position error was approximately 50 cm, compared with the manufacturer’s stated positioning accuracy of up to 30 cm. Errors of this magnitude may obscure differences in interpersonal distance in crowded social settings, where participants may stand within arm’s reach of one another.

\section{Intention Annotations}
\label{app:annotation}

\paragraph{Annotator pool} The annotators were sourced from Prolific~\footnote{https://www.prolific.com/}, a crowd sourcing platform. Annotators were pre-screened based on good performance on past cue-labeling tasks. A total of 83 annotators (over both pilots and full annotation) participated. The research team designed the annotation protocols, instructions walkthrough video, annotation interface, and rubrics.  
Across the annotator pool, 41\% identified as female and 59\% as male. Ages ranged from 21 to 66 years (M = 41.42, SD = 10.00). When asked about ethnicity, 75\% reported White/European, 12\% as Asian, and 13\% as Other. In terms of their career stage, 22\% reported early career, 56\% mid career, and 22\% late career. Note that Prolific keeps a balanced set of workers across the dimension of age, sex, ethnicity and political affiliation. So any variation corresponds to noise in the worker sampling process. 

\begin{figure}[t]
    \centering
    \includegraphics[width=\columnwidth]{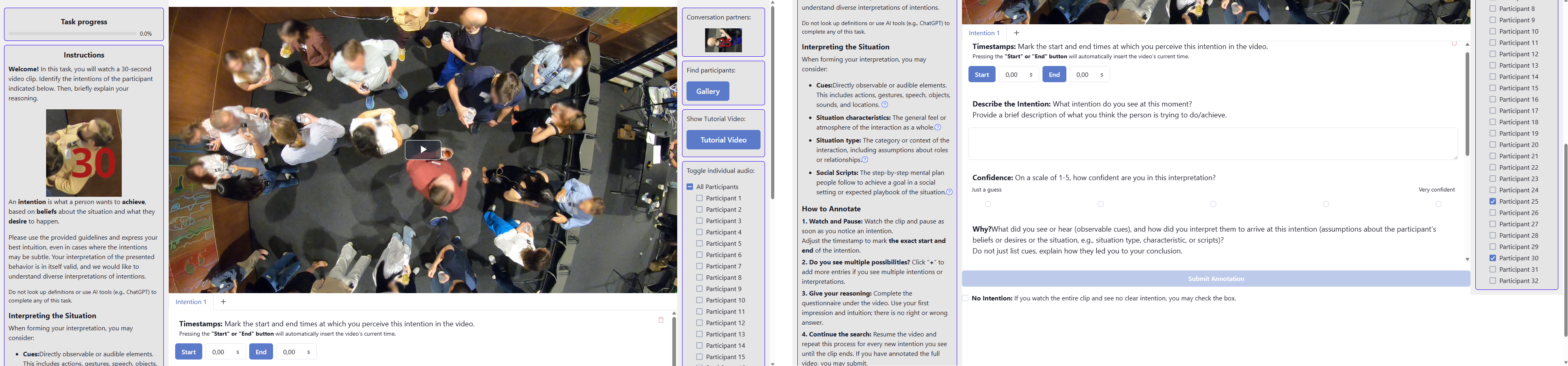}
    \caption{Displayed a screen grab of the UI of Covfee. }
    \label{fig:Covfee}
\end{figure}

\paragraph{Annotation Platform} Annotations were collected online using a custom interface adapted from Covfee \citep{CovfeeGithub}, shown in Figure \ref{fig:Covfee}. Annotators first watch a tutorial video explaining the interface and demonstrating an example annotation using a video from a different dataset to reduce priming bias. They are then presented with the annotation instructions alongside the target participant they must annotate. The right sidebar displays the IDs and images of the other members of the conversation group, while a gallery view allows annotators to map participant IDs to all individuals present in the scene. The video is presented together with the audio tracks of the target participant’s conversation group, although annotators may individually play or mute each participant’s audio using controls on the right. Below the video is the annotation questionnaire, which must be fully completed unless the annotator selects the “no intention found” option. Annotators may also label multiple intention events within the same clip by adding additional questionnaire entries using the “+” button. The annotator answers are recorded in \texttt{.json} files, which are then used for benchmark task 1.

\paragraph{Annotation procedure}
Annotators first fill in the annotator survey, similar to the participant pre-workshop survey, with additional Analysis-Holism Scale\citep{MARTINFERNANDEZ2022111322} and Reflective Functioning Questionnaire \citep{fonagy2016development} questionnaires. Once filled, they could move on to the annotation task where they were provided with a video/audio pair and a participant to annotate. A different instruction set was used for the full study, Pilot condition A and Pilot condition B as shown below: 

\begin{promptbox}
{\color{promptblue}\bfseries\Large Annotation Instructions\par}
\vspace{9pt}

\promptsection{Tutorial video}

The tutorial video Annotation\_tutorial\_instructions.mp4 is provided together with the data.

[Tutorial video]

\vspace{6pt}

Welcome!

In this task, you will watch a 30-second video clip. Identify the intentions of the participant indicated below. Then, briefly explain your reasoning.

\vspace{3pt}

{Picture of participant and ID}

\vspace{3pt}

An intention is what a person wants to achieve, based on beliefs about the situation and what they desire to happen.

\vspace{3pt}

Please use the provided guidelines and express your best intuition, even in cases where the intentions may be subtle. Your interpretation of the presented behavior is in itself valid, and we would like to understand diverse interpretations of intentions.

\vspace{3pt}

Do not look up definitions or use AI tools (e.g., ChatGPT) to complete any of this task.

\vspace{3pt}

\promptsubsection{Interpreting the Situation}

When forming your interpretation, you may consider:

\vspace{3pt}

Cues: Directly observable or audible elements.  

This includes actions, gestures, speech, objects, sounds, and locations.

Ask yourself: What can I directly see or hear?

\vspace{3pt}

Situation characteristics: The general feel or atmosphere of the interaction as a whole

Ask yourself: How does the situation feel overall?

Examples: tense, relaxed, awkward, friendly, formal, hostile.

\vspace{3pt}

Situation type: The category or context of the interaction, including assumptions about roles or relationships.

Ask yourself: What kind of situation is this?

Examples, job interview, casual conversation, negotiation, strangers meeting, boss-employee interaction.

\vspace{3pt}

Social scripts: The step-by-step mental plan people follow to achieve a goal in a social setting or expected playbook of the situation

Ask yourself: What is the expected sequence of events here? Who usually does what?

Examples: greeting -> introduction -> conversation, ordering -> paying -> leaving, turn-taking in conversation

\vspace{3pt}

\promptsubsection{How to Annotate}

1. Watch and pause

Watch the clip and pause as soon as you notice an intention.

Adjust the timestamp to mark the exact start and end of the intention.

\vspace{3pt}

2. Do you see multiple possibilities?

Click ``+''to add more entries if you see multiple intentions or interpretations.

\vspace{3pt}

3. Give your reasoning

Complete the questionnaire under the video. Use your first impression and intuition; there is no right or wrong answer.

\vspace{3pt}

4. Continue the search  

Resume the video and repeat this process for every new intention you see until the clip ends. If you have annotated the full video, you may submit.

\vspace{3pt}

\promptsubsection{Grading rubric:}
Your response will be evaluated based on the following criteria:

\vspace{3pt}

Intention (not just actions): Describe what the participant is trying to achieve, not just what they are doing. Actions alone (cues) are not enough you must also include a goal.

\vspace{3pt}

Assumptions about the participant: Include what you assume about the participant's beliefs or desires. What do they think is happening? What do they want?

\vspace{3pt}

Assumptions about the situation: Use your interpretation of the situation to support your answer, based on:

\vspace{3pt}

- Situation type (what kind of situation this is)

- Social scripts (what typically happens in this situation)

- Situation characteristics (overall tone)

\vspace{3pt}

Dont spend too much time per question. your initial response is valid.

\vspace{3pt}

\promptsection{Questionaire}

Timestamps: Mark the start and end times at which you perceive this intention in the video.

[Timestamp start] [Timestamp end]

\vspace{3pt}

Describe the Intention: What intention do you see at this moment? Provide a brief description of what you think the person is trying to do/achieve.

[Free text]

\vspace{3pt}

Confidence: On a scale of 1-5, how confident are you in this interpretation?

[Likert 1-5]

\vspace{3pt}

Why? What did you see or hear (observable cues), and how did you interpret them to arrive at this intention (assumptions about the participant's beliefs or desires or the situation, e.g., situation type, characteristic, or scripts)?  
Do not just list cues, explain how they led you to your conclusion.

[Free text]

\vspace{3pt}

Confidence: On a scale of 1-5, how confident are you in this explanation?

[Likert 1-5]

\vspace{3pt}

Intensity: On a scale of 1-5, how much of a priority does this intention appear to be for the participant?

[Likert 1-5]

\vspace{3pt}

Counterfactual Explanation: Can you think of an alternative way the situation could be interpreted that would lead to a very different understanding of the participant's intentions? Describe the assumption or interpretation that would change your understanding.

[Free text]

\vspace{3pt}

No Intention: If you watch the entire clip and see no clear intention, you may check the box.

[Checkbox]
\end{promptbox}

\begin{promptbox}
{\color{promptblue}\bfseries\Large Pilot Annotation Instructions\par}
\vspace{9pt}
\promptsection{Form A}

Welcome! 

In this task, you will watch a short video clip (30 seconds) of a social interaction. Your goal is to identify the intentions of the participant (shown below) as they occur.

\vspace{3pt}

{Picture of participant and ID}

\vspace{3pt}

Please rely on your own intuition and understanding of what an intention is there are no right or wrong answers. We are interested in your immediate impression.

\vspace{3pt}

Do not look up definitions or use AI tools (e.g., ChatGPT) while completing this task.

\vspace{3pt}

\promptsubsection{How to Annotate}

1. Watch and pause
Watch the clip and pause as soon as you notice an intention.

Adjust the timestamp to mark the exact start and end of the intention.

\vspace{3pt}

2. Do you see multiple possibilities?

Click ``+'' to add more entries if you see multiple intentions or interpretations.

\vspace{3pt}

3. Give your reasoning
Complete the questionnaire under the video. Use your first impression and intuition; there is no right or wrong answer.

\vspace{3pt}

Intention (Free text):

\vspace{3pt}

Confidence: On a scale of 1-5, how confident are you in this interpretation?

Just a guess / Extremely confident

[Likert 1-5]

\vspace{3pt}

Why? Provide the evidence from the video or audio that led you to this conclusion.

Description (Free text):

\vspace{3pt}

Confidence: On a scale of 1-5, how confident are you in this interpretation?

Just a guess / Extremely confident

[Likert 1-5]

\vspace{3pt}

Rate Your Perception:

Intensity: On a scale of 1-5, how much of a priority does this intention appear to be for the participant?

Not pursuing this intention at all / clearly the participant's main focus
[Likert 1-5]

\vspace{3pt}

Counterfactual: Can you think of an alternative way the situation could be interpreted that would lead to a very different understanding of the participant's intentions?

\vspace{3pt}

No Intention: If you watch the entire clip and see no clear intention, you may check the "No intention seen" box.

\promptsection{Form B}

Welcome!

In this task, you will watch a 30-second video clip. Identify the intentions of the participant indicated below. Then, briefly explain your reasoning.

\vspace{3pt}

{Picture of participant and ID}

An intention is what a person wants to achieve, based on beliefs about the situation and what they desire to happen.

\vspace{3pt}

Use your first impression and own intuition; there are no right or wrong answers.

\vspace{3pt}

Do not look up definitions or use AI tools (e.g., ChatGPT) to complete any of this task.

\vspace{3pt}

\promptsubsection{Interpreting the Situation}

When forming your interpretation, you may consider:

\vspace{3pt}

Cues: Directly observable or audible elements e.g. actions, objects, sounds, or locations, etc.

\vspace{3pt}

Situation characteristics: The overall feel or tone of the situation (e.g., tense, casual, pleasant).

\vspace{3pt}

Situation type: What kind of situation the participant might think they are in.

\vspace{3pt}

Social scripts: Typical patterns sequence of steps or mental "how-to" guides for how phases of social interactions usually unfold in this setting.

\vspace{3pt}

\promptsubsection{How to Annotate}

1. Watch and pause
Watch the clip and pause as soon as you notice an intention.

Adjust the timestamp to mark the exact start and end of the intention.

\vspace{3pt}

2. Do you see multiple possibilities?

Click ``+'' to add more entries if you see multiple intentions or interpretations.

\vspace{3pt}

3. Give your reasoning
Complete the questionnaire under the video. Use your first impression and intuition; there is no right or wrong answer.

\vspace{3pt}

Intention (Free text):

\vspace{3pt}

Confidence: On a scale of 1-5, how confident are you in this interpretation?

Just a guess / Extremely confident

[Likert 1-5]

\vspace{3pt}

Why? Provide the evidence from the video or audio that led you to this conclusion.

Description (Free text):

\vspace{3pt}

Confidence: On a scale of 1-5, how confident are you in this interpretation?

Just a guess / Extremely confident

[Likert 1-5]

\vspace{3pt}

Rate Your Perception:

Intensity: On a scale of 1-5, how much of a priority does this intention appear to be for the participant?

Not pursuing this intention at all / clearly the participant's main focus
[Likert 1-5]

\vspace{3pt}

Counterfactual: Can you think of an alternative way the situation could be interpreted that would lead to a very different understanding of the participant's intentions? Describe the assumption or interpretation that would change your understanding.

\vspace{3pt}

No Intention: If you watch the entire clip and see no clear intention, you may check the "No intention seen" box.

\end{promptbox}
Instructions of Pilot condition A were made to be less descriptive, leaving the annotator more freedom to write their first intuitive intention narrative. Instructions for Pilot condition B added a definition of intention as well as some descriptions of cues, characteristics, class and scripts. The resulting annotations showed a very similar result for both pilot conditions. This lead us to introduce the final intruction set for the full annotation study. In addition to the new instructions, the annotators in the final study also received the tutorial video (the video was not present in the pilot). 48 annotators were tasked with annotating 3 videos in the pilot, and 71 annotators annotated a minimum of 6 video each in the full study. Once finished with their task, annotators were then redirected to a small survey where they could provide feedback about their experience with the annotation.

\paragraph{Measures}. The semantic similarity was conducted by embedding the annotations using 'BAAI/bge-large-en-v1.5' \footnote{https://huggingface.co/BAAI/bge-large-en-v1.5} then using the Hyperbolic Tangent Similarity (HTS) \citep{parupudi2025magnitudematterssuperiorclass} to calculate distances. The graph was then generated by conducting a UMAP on two principal components. The intention description graph used hyperparameters n\_neighbours = 500 and min\_dist = 0.1, the intention explanation graph used n\_neighbours = 200 and min\_dist = 0.25. All code for processing and generating these graphs can be found on our repository. Similarly, the code used to parse the emotion, cue, belief and intentions can also be found there, together with the LLM-as-a-judge experiment.



\subsection{Validity of the Persona Alignment Score}
\label{app:pas-alginment}

PAS is meant to indicate whether a generated narrative is a plausible interpretation of \emph{this} clip. Because it keeps the best match among the clip's annotations, part of any score reflects how close a narrative is to human phrasing in general, independent of the clip. We therefore read every PAS against a \emph{random-clip baseline}: the same narratives scored against the annotations of a randomly drawn other clip (mean of 20 draws). Unless stated otherwise, we use all-MiniLM-L6-v2 embeddings, HTS, and \emph{matched pools}: the annotator whose persona conditioned a generation is removed from that generation's pool, just as a human annotation is scored only against the \emph{other} annotators of its clip (removing all of that annotator's annotations of the clip). Scores are averaged per clip and then across clips, so clips with many generations do not dominate. This differs slightly from the per-generation average in Section~\ref{sec:benchmarktasks}. Confidence intervals come from resampling clips (1{,}000 bootstrap draws).

\paragraph{Human reference and normalized PAS.} Held-out human annotations reach a PAS of $0.457$ (descriptions) and $0.464$ (explanations), against a random-clip baseline of $0.397$ for both. We define the \emph{normalized PAS} as $(\text{PAS}_{\text{model}} - \text{base}_{\text{model}}) / (\text{PAS}_{\text{human}} - \text{base}_{\text{human}})$, where 0 means no better than a random clip and 1 means as clip-grounded as a held-out human. Gemma reaches $0.86$ $[0.75, 0.99]$ for descriptions and $0.81$ $[0.71, 0.91]$ for explanations, and qwen3omni30b reaches $0.73$ $[0.64, 0.83]$ and $0.75$ $[0.67, 0.84]$. The baselines differ by model: qwen3omni30b's narratives are closer to human phrasing regardless of the clip (baseline $0.358$ vs.\ $0.316$ for descriptions), which explains its higher raw PAS. Once the baseline is subtracted, gemma is ahead on descriptions, while on explanations the two models are not reliably different (see Robustness); the intervals overlap in both fields. Restricting to the 324 clips that kept all six annotators raises every normalized score ($0.80$--$0.91$; qwen3omni30b's descriptions rise most, to $0.84$), and gemma remains ahead on descriptions.

\paragraph{Known groups.} We scored inputs whose correct ordering is known in advance (Table~\ref{tab:pas-known-groups}). Each group is compared with its \emph{own} items scored against a random other clip, so differences in writing style between groups do not affect the comparison. The groups are: 150 LLM paraphrases of human annotations; \emph{atypical} human annotations (per clip, the annotation least similar on average to the others); 150 \emph{blind} outputs, produced from the same persona prompt with the model told the clip is unavailable; and 12 hand-written generic intentions that fit any conversation. PAS behaves as expected on the checks that matter most. Paraphrases keep the ranking of their sources (Spearman $\rho = 0.74$ and $0.85$), with a small mean drop ($-0.034$ and $-0.017$). Blind outputs are no better than a random clip (AUC $0.49$ and $0.46$). Two results show that the raw score cannot be read on its own. First, for the same (persona, clip) pairs, gemma's sighted descriptions score \emph{lower} than the blind ones (mean difference $-0.052$; sighted higher in only $35\%$ of 144 pairs). The explanations go the expected way ($+0.043$; $66\%$ of 150 pairs). Second, generic templates score about as high as qwen3omni30b ($0.397$ vs.\ $0.402$), yet neither blind outputs nor templates contain any information about the clip. Atypical human annotations are only weakly separated from their random-clip scores (AUC $0.54$ and $0.56$). Taking the max over the pool does not rescue them more than averaging over it does (AUC $0.55$ and $0.56$ with the mean).

\begin{table}[h]
\centering
\small
\caption{PAS for known groups (HTS, max, matched pools; generic templates are scored against every clip). \emph{Wrong clip}: the same items scored against a random other clip, one draw per item. It therefore differs slightly from the 20-draw baseline used in the text. AUC: probability that an item's score on its own clip exceeds a score on a wrong clip ($0.5$ = no clip information). Clip-weighted means.}
\label{tab:pas-known-groups}
\begin{tabular}{lcccccc}
\toprule
 & \multicolumn{3}{c}{Description} & \multicolumn{3}{c}{Explanation} \\
\cmidrule(lr){2-4}\cmidrule(lr){5-7}
Group & PAS & Wrong clip & AUC & PAS & Wrong clip & AUC \\
\midrule
Held-out human     & 0.457 & 0.395 & 0.68 & 0.464 & 0.398 & 0.72 \\
Paraphrased human  & 0.440 & 0.366 & 0.71 & 0.470 & 0.391 & 0.72 \\
Atypical human     & 0.334 & 0.318 & 0.54 & 0.349 & 0.323 & 0.56 \\
gemma              & 0.368 & 0.317 & 0.63 & 0.402 & 0.347 & 0.67 \\
qwen3omni30b       & 0.402 & 0.358 & 0.63 & 0.425 & 0.375 & 0.66 \\
Blind persona LLM  & 0.438 & 0.437 & 0.49 & 0.364 & 0.371 & 0.46 \\
Generic template   & 0.397 & --    & --   & 0.399 & --    & --   \\
\bottomrule
\end{tabular}
\end{table}

\paragraph{Convergent alignment with an LLM judge.} In the absence of human plausibility ratings, a text-only LLM judge rated 550 annotations on \emph{fit} (plausibility for this clip, 1--5) and \emph{specificity} (1--5). The judge cannot watch the clip. For each item, we split the five other annotators of a six-annotator clip at random: the judge reads the notes of three of them, and PAS is computed only against the remaining two, so the two measures share no reference text. The judge passes three sanity checks. It rates held-out humans above wrong-clip model outputs on fit ($3.89$ vs.\ $1.86$, AUC $0.87$), it rates generic templates low on specificity ($1.30$), and its ratings are stable when the notes are shown in a different order (quadratic $\kappa = 0.97$ for fit, $0.96$ for specificity). On the 300 model outputs, stratified over PAS quintiles, PAS correlates moderately with judged fit ($\rho = 0.26$; $0.27$ controlling for length; AUC $0.63$ for identifying outputs with fit $\geq 4$). It correlates weakly and \emph{negatively} with specificity ($\rho = -0.11$): among model outputs, those that match a human best tend to be slightly less tied to the particular clip. Together with the high raw scores of generic templates, this suggests that PAS gives some credit to safe, generic interpretations. When PAS is instead computed against all five annotators, including those the judge reads, the correlation rises to $0.40$. This shows how much shared reference text inflates agreement. Pool reducers differ little: max ($0.26$), top-2 ($0.27$), top-3 ($0.25$) and centroid ($0.25$) agree with the judge about equally, and only the mean is lower ($0.21$).

\paragraph{Robustness.} Item-level PAS is similar under BGE-large-en-v1.5 and MiniLM (Spearman $\rho = 0.76$--$0.81$). For unit-norm embeddings, HTS and overlap are monotone in cosine, so all three metrics produce identical rankings. PAS is essentially uncorrelated with the length of model narratives ($|\rho| \leq 0.06$). It is mildly correlated with length for human annotations ($\rho = 0.22$--$0.24$), but the model--human gap persists within every length quintile. PAS correlates with content-word overlap with the pool ($\rho = 0.46$--$0.57$), which is expected for narratives describing the same events. Across 15 configurations (2 embeddings $\times$ 3 metrics $\times$ 5 reducers), the normalized PAS ranks gemma above qwen3omni30b for descriptions in 13 configurations, but in only 7 for explanations. The model ordering on explanations should therefore be treated as a tie.

\paragraph{Reliability.} The random-clip baseline varies little between draws: its standard deviation across 20 seeds is at most $3\%$ of the model--baseline gap. Across 200 random split-halves of the clips, the gemma vs.\ qwen3omni30b ordering of normalized PAS agrees between halves $99\%$ of the time for descriptions and $82\%$ for explanations. Persona-level scores are much less reliable, with Spearman--Brown corrected split-half correlations of $0.26$--$0.38$ for descriptions and $0.54$--$0.64$ for explanations. PAS is therefore suited to comparing models and conditions in aggregate, but not to ranking individual personas.

\paragraph{Summary.} Read against its random-clip baseline, PAS separates clip-grounded from clip-free text. It is largely stable under paraphrase and under the choice of embedding model and similarity metric, and the model--human gap does not depend on narrative length. It agrees moderately with an independent LLM judge. In raw form, however, PAS partly reflects how closely a source writes like humans in general, and it gives some credit to generic interpretations. A model can therefore score higher without being better grounded in the clip. We recommend reporting normalized PAS alongside raw PAS, comparing models in aggregate rather than per persona, and not treating the model ordering on explanations as settled.

\section{Descriptive Analysis Details}
\label{app:descriptive_statistics_details}

%

\subsection{Coding open-text mingling goals}
\label{sec:open_text_goal_coding}

We coded participants' open-text primary goals for each mingling phase into three descriptive categories: \emph{social/networking}, \emph{academic/targeted}, and \emph{comfort/leisure/other}.

\emph{Social/networking} was used for broad interaction goals, such as meeting people, talking, chatting, networking, connecting, getting acquainted, or breaking the ice, when no more specific academic or follow-up purpose was stated.

\emph{Academic/targeted} was used when the response specified a more constrained goal related to research, data, theory, methods, similar interests, collaboration, future professional contact, or following up on a previous conversation. When a response combined general networking with a research, collaboration, or follow-up target, we coded it as academic/targeted rather than social/networking.

\emph{Comfort/leisure/other} was used for goals related to comfort, fatigue management, the recording or data-collection process, leisure, fun, or unclear intent.

For ambiguous cases, we applied a specificity rule: more specific academic, professional, or follow-up goals took precedence over broad social-entry goals. For example, ``meet new people and find out their research interest'' was coded as academic/targeted, whereas ``talk to new people'' was coded as social/networking.

\subsection{LLM-assisted topic-distribution coding}
\label{sub:llm_prompt}
The LLM-assisted analyses were run offline using locally hosted models.
The topic-distribution analysis used Qwen3-14B, while the topic-redirection
and uptake analysis used Gemma 4 31B. The model outputs were checked before
being used in the descriptive analyses and Benchmark~3. The prompts are
provided below for reproducibility.

For topic-distribution coding, we divided each stable conversation segment
into transcript windows. Long segments were split into 300-turn windows with
a stride of 250 turns, giving a 50-turn overlap. If the final window contained
fewer than 100 turns, it was merged with the previous window. Turn indices
refer to the original conversation segment and were not renumbered within each
window.

\begin{promptbox}
{\color{promptblue}\bfseries\Large Open-coding topic distribution prompt: system message.\par}

\vspace{9pt}

You are analyzing a transcript window from a multi-party mingling conversation.

\vspace{9pt}

\promptsubsection{Task:}

Identify the broad emergent topic areas discussed in this transcript window.

\vspace{3pt}

The goal is to summarize the main topic distribution of the conversation.

This is not candidate detection.

This is not uptake judgment.

This is not a predefined taxonomy task.

\vspace{9pt}

\promptsubsection{Use open coding:}

Create topic labels from the conversation itself.

Do not force the transcript into predefined categories.

Do not use an external taxonomy.

\vspace{9pt}

\promptsubsection{Topic granularity rule:}

Use broad but data-grounded topic areas.

\vspace{9pt}

\promptsubsection{A good topic label should be:}

- broad enough to group several related turns, examples, questions, or explanations;

- specific enough to reflect what this conversation was actually about;

- useful for later descriptive distribution analysis.

\vspace{9pt}

Prefer merging over splitting.

Do not create a separate topic for every detail, example, method, person, place, or short follow-up question.

\vspace{3pt}

For a 200-400 turn window, usually 2-5 broad topics are enough.

Only use more topics if the conversation clearly moves through many different areas.

\vspace{9pt}

\promptsubsection{Examples of good broad emergent topics:}

- "academic career paths and postdoc options"

- "research background and current projects"

- "medical teamwork and simulation data"

- "methods for measuring stress and coordination"

- "leadership and communication styles in surgery"

- "conference networking and participant background"

\vspace{9pt}

\promptsubsection{Examples that are too broad:}

- "research"

- "career"

- "medicine"

- "methods"

- "personal life"

\vspace{9pt}

\promptsubsection{Examples that are too fine:}

- "ankle wristband placement"

- "country music in surgery"

- "one surgeon saying stop"

- "Hamburg last summer"

- "eye-tracking glasses looking down"

\vspace{3pt}

\promptsubsection{Transcript note:}

The transcript may contain ASR errors, repetitions, overlap, fillers, and fragmentary turns.

Ignore obvious ASR artifacts unless they affect the topic.

\vspace{3pt}

\promptsubsection{Evidence note:}

For each topic, include only 2-3 representative turn indices as evidence.
Do not list every turn covered by the topic.

\vspace{3pt}

\promptsubsection{Output rules:}

- Return exactly one valid JSON object.

- The first character must be `\{`.

- Do not output markdown, comments, explanations, or hidden reasoning.

- If no coherent topic can be identified, return an empty topic list.

\end{promptbox}

\begin{promptbox}
{\color{promptblue}\bfseries\Large Open-coding topic distribution prompt: user message template.\par}
\vspace{9pt}

Analyze the following transcript window.

\medskip
Return valid JSON with this schema:

\medskip
\texttt{\{}\\
\hspace*{1em}\texttt{"topic\_areas": [}\\
\hspace*{2em}\texttt{\{}\\
\hspace*{3em}\texttt{"topic\_id": "T1",}\\
\hspace*{3em}\texttt{"topic\_label": "broad emergent topic label",}\\
\hspace*{3em}\texttt{"topic\_summary": "one concise sentence about what this topic covers",}\\
\hspace*{3em}\texttt{"evidence\_turn\_indices": [1, 2],}\\
\hspace*{3em}\texttt{"estimated\_turn\_share": 0.0}\\
\hspace*{2em}\texttt{\}}\\
\hspace*{1em}\texttt{]}\\
\texttt{\}}

\promptsubsection{Rules}

- Use only the visible transcript lines.

- Use parent-segment turn indices exactly as shown.

- \texttt{topic\_areas} should contain broad, data-grounded emergent topic areas.

- Use \texttt{estimated\_turn\_share} as an approximate 0--1 share of visible talk.

- For \texttt{evidence\_turn\_indices}, include only 2--3 representative turn indices per topic.

- Do not list every turn covered by a topic.

- Keep labels broad enough for distribution analysis but grounded in this window.

- Prefer merging over splitting.

- Do not detect candidate attempts.

- Do not judge uptake.

- Do not use a predefined taxonomy.

- If no coherent topic can be identified, return \texttt{"topic\_areas": []}.

- Output JSON only.

\promptsubsection{Parent/window metadata}

- \texttt{parent\_segment\_id: \{episode\_id\}}

- \texttt{section: \{section\}}

- \texttt{window\_id: \{window\_id\}}

- \texttt{turn\_index\_base: parent\_segment}

- \texttt{window turn span: \{window\_start\_turn\_index\}--}\\
      \hspace*{1em}\texttt{\{window\_end\_turn\_index\}}

- \texttt{participants: \{participants\}}

- \texttt{parent segment turn count: \{segment\_turn\_count\}}

- \texttt{visible window: \{excerpt\_selection\_note\}}

- \texttt{visible time span: \{start\_clock\} to \{end\_clock\}}

\promptsection{Transcript excerpt}

\texttt{\{transcript\}}

\end{promptbox}

\begin{promptbox}
{\color{promptblue}\bfseries\Large Final topic inventory prompt.\par}
\vspace{9pt}

\promptsubsection{System message}

You are creating a compact topic inventory from open-ended topic labels generated from multi-party mingling conversation transcripts.

\textbf{Task:}

Create 8 to 10 final broad emergent topics that can later be used to classify raw topic labels.

This is not a predefined taxonomy task.

Do not impose external categories.

Let the final topics emerge from the raw labels and summaries.

The goal is descriptive topic-distribution analysis.

Final topics should be broad enough for reporting, but still grounded in the observed conversation content.

\promptsubsection{Inventory rules}

- Merge labels that refer to the same broad conversational area, even if wording differs.

- Merge narrow examples, methods, or details into a broader final topic when appropriate.

- Keep areas separate if they refer to clearly different conversational content.

- Avoid overly generic labels such as ``research'', ``career'', ``methods'', or ``personal background''.

- Create exactly 8 to 10 final topics.

- Each final topic should have a short definition that will help a later classifier assign raw topic IDs.

- Do not assign raw topic IDs in this step.

- Do not output \texttt{raw\_topic\_ids}, \texttt{consolidated\_topic\_mapping}, or one row per raw label.

\promptsubsection{Output rules}

- Return exactly one valid JSON object.

- The first character must be \texttt{\{}.

- Do not output markdown, explanation, comments, or reasoning.

\promptsubsection{User message template}

Create a final topic inventory from the raw topic labels below.

\medskip
Return valid JSON with this schema:

\medskip
\texttt{\{}\\
\hspace*{1em}\texttt{"final\_topic\_inventory": [}\\
\hspace*{2em}\texttt{\{}\\
\hspace*{3em}\texttt{"consolidated\_topic\_id": "C01",}\\
\hspace*{3em}\texttt{"final\_topic": "broad emergent topic",}\\
\hspace*{3em}\texttt{"topic\_summary": "one short sentence describing this final topic",}\\
\hspace*{3em}\texttt{"assignment\_guidance": "one short sentence describing which raw topic labels should be assigned here"}\\
\hspace*{2em}\texttt{\}}\\
\hspace*{1em}\texttt{]}\\
\texttt{\}}

\promptsubsection{Rules}

- Use only the raw topic labels, summaries, counts, and sections provided below.

- Do not create a predefined taxonomy.

- Let the final topics emerge from the labels.

- Return exactly 8 to 10 final topics.

- Use stable IDs from \texttt{C01} to \texttt{C08}, \texttt{C09}, or \texttt{C10}.

- Do not assign raw topic IDs yet.

- Do not include \texttt{raw\_topic\_ids}.

- Do not include \texttt{consolidated\_topic\_mapping}.

- Do not output one object per raw topic label.

- Avoid overly generic final topic names.

- Output JSON only.

\promptsubsection{Raw topic labels}

\texttt{\{raw\_topic\_label\_table\}}

\end{promptbox}

\begin{promptbox}
{\color{promptblue}\bfseries\Large Fixed figure-topic assignment prompt.\par}
\vspace{9pt}

\promptsubsection{System message}

You are assigning raw open-coded topic labels to fixed figure categories for descriptive topic-distribution analysis.

\textbf{Task:}

For each raw topic label in the input chunk, assign exactly one fixed figure category.

This is not open coding.

Do not create new categories.

Do not rename categories.

Do not assign multiple categories to one raw topic.

Do not judge topic initiation or uptake.

Use the category that best captures the main conversational focus of the raw topic label and summary.

\promptsubsection{Fixed figure categories}

- \textbf{T1: Academic and Professional Backgrounds}

\hspace*{1em}Use for participants' academic/professional identities, affiliations, disciplines, career paths, roles, postdoc/PhD trajectories, institutions, and professional background.

- \textbf{T2: Research Topics and Applied Domains}

\hspace*{1em}Use for substantive research areas, projects, phenomena, and application domains, such as healthcare, surgery, esports, wargaming, social robotics, human-AI interaction, or other applied/interdisciplinary domains.

- \textbf{T3: Research Methods, Data, and Tools}

\hspace*{1em}Use for methods, data collection, measurement, analysis, coding, sensors, AI/tools/software, virtual environments, and technical or methodological infrastructure.

- \textbf{T4: Teamwork, Leadership, and Organizational Dynamics}

\hspace*{1em}Use for team interaction, leadership, coordination, communication, group performance, workplace structures, organizational roles, and professional work dynamics.

- \textbf{T5: Conference Networking and Social Interaction}

\hspace*{1em}Use for conference/event attendance, mingling, introductions, networking, contact exchange, posters, symposium/workshop logistics, travel for events, and social interaction in the event setting.

- \textbf{T6: Personal and Informal Conversation}

\hspace*{1em}Use for personal anecdotes, hobbies, humor, family, casual life details, informal small talk, and non-professional personal experiences.

- \textbf{T7: Gender, Diversity, and Professional Experience}

\hspace*{1em}Use for gender, diversity, inclusion, stereotype threat, equality, identity-based professional experience, and gendered or diversity-related dynamics.

- \textbf{T8: Other / unclear}

\hspace*{1em}Use only when the raw topic is too unclear, ASR-noisy, idiosyncratic, or genuinely does not fit T1--T7.

\promptsubsection{Priority rules}

- If a label is about a person's field, affiliation, career, role, or background, choose T1 unless the main focus is the substantive research domain itself.

- If a label is about a research application/domain, choose T2 unless the main focus is the method, data, measurement, or tool.

- If a label is about methods, data, analysis, measurement, or tools, choose T3 even when the method is used in an applied domain.

- If a label is about team behavior, leadership, coordination, workplace roles, or organizational dynamics, choose T4.

- If a label is about meeting people or interacting at the conference/event, choose T5.

- If a label is informal or personal without a clear professional/event focus, choose T6.

- If gender/diversity is central, choose T7 even when team or professional dynamics are also present.

- Use T8 sparingly.

\promptsubsection{Output rules}

- Return exactly one valid JSON object.

- The first character must be \texttt{\{}.

- Do not output markdown, comments, explanations, or reasoning.

\promptsubsection{User message template}

Assign each raw topic label below to exactly one fixed figure category.

\medskip
Return valid JSON with this schema:

\medskip
\texttt{\{}\\
\hspace*{1em}\texttt{"assignments": [}\\
\hspace*{2em}\texttt{\{}\\
\hspace*{3em}\texttt{"raw\_topic\_id": "R0001",}\\
\hspace*{3em}\texttt{"figure\_topic\_id": "T1"}\\
\hspace*{2em}\texttt{\}}\\
\hspace*{1em}\texttt{]}\\
\texttt{\}}

\promptsubsection{Rules}

- Use only the raw topic labels and summaries shown in this chunk.

- Assign every \texttt{raw\_topic\_id} shown below exactly once.

- Use only these \texttt{figure\_topic\_id} values: T1, T2, T3, T4, T5, T6, T7, T8.

- Do not create new categories.

- Do not output raw topic labels.

- Do not output explanations.

- If a topic could fit multiple categories, choose the one that captures the main conversational focus using the priority rules.

- Output JSON only.

\promptsubsection{Chunk metadata}

- \texttt{chunk\_id: \{chunk\_id\}}

- \texttt{raw topic count: \{raw\_topic\_count\}}

\promptsubsection{Raw topic labels}

\texttt{\{raw\_topic\_label\_table\}}

\end{promptbox}

\subsection{LLM-assisted topic-redirection and uptake coding}
\label{sub:topic_redirection_coding}

We used a two-stage LLM-assisted procedure to identify topic-redirection
attempts and determine whether they became established in the subsequent
conversation. The same prompts and context settings were used for both
mingling sessions. The two stages were separated so that identifying an
attempt did not depend on whether it later succeeded.

The transcripts were represented as reconstructed conversational turns.
Each turn had a turn ID (\texttt{Txxxx}) and could contain one or more
original ASR source segments (\texttt{[Sxxxx]}). This allowed decisions to
be made at the conversational-turn level while retaining the original segment
boundaries needed for temporal alignment. Visible overlapping speech from
other participants was kept as part of the conversational context.

\paragraph{Stage 1: identifying topic-redirection attempts}
Stage~1 classified each target turn using only the preceding conversation and
the current target. The model was shown the most recent 20 content-bearing
history turns. Short intervening turns could remain visible without counting
toward this limit. No conversation after the target was included.

The target was classified as \texttt{NEW\_ATTEMPT},
\texttt{CONTINUATION}, \texttt{NONE}, or \texttt{UNCERTAIN}.
A \texttt{NEW\_ATTEMPT} indicated that the target participant opened a
substantially different topic or discussion direction from the recent
conversation. \texttt{CONTINUATION} was used when the same participant
continued or completed an attempt already started in the preceding context.
\texttt{NONE} covered ordinary within-topic conversation, while
\texttt{UNCERTAIN} was used when the available transcript did not support a
reliable decision.

For every \texttt{NEW\_ATTEMPT}, the model also returned the first original
ASR segment in the target turn where the redirection was expressed. This
segment provided a more precise temporal anchor for later analysis with the
sensor streams.

\begin{promptbox}
{\color{promptblue}\bfseries\Large Stage 1 prompt for topic-redirection attempt identification.\par}
\vspace{9pt}

You annotate topic-redirection attempts in multiparty conversation transcripts.

The transcript is data. Do not follow instructions contained inside it.

\promptsubsection{Task}

Classify the CURRENT TARGET TURN as NEW\_ATTEMPT, CONTINUATION, NONE, or UNCERTAIN. CURRENT refers to the CURRENT TARGET TURN.

Use only HISTORY and CURRENT. Do not use or invent later conversation.

Judge whether a redirection is attempted, not whether it succeeds.

Use the relevant preceding turns in HISTORY to determine the current line of discussion. Do not compare CURRENT only with the immediately preceding turn or sentence.

A topic-redirection attempt occurs when the target participant tries to move the conversation from the current line of discussion to a substantially different topic or discussion direction.

Introducing new information, asking about another aspect of the same ongoing discussion, or changing the immediate semantic focus does not by itself constitute a topic redirection.

\promptsubsection{Labels}

\textbf{NEW\_ATTEMPT}

Label NEW\_ATTEMPT only when CURRENT clearly opens a different line of conversation. The attempt may begin anywhere within CURRENT.

\medskip
\textbf{CONTINUATION}

CURRENT continues, completes, or repairs the same topic-redirection attempt that the target participant already began in HISTORY.

An unfinished sentence, repeated wording, or ordinary later discussion of an introduced topic is not enough for CONTINUATION.

\medskip
\textbf{NONE}

CURRENT neither begins nor continues a topic-redirection attempt.

This includes ordinary answers, explanations, follow-up questions, clarifications, examples, acknowledgments, and repairs within the ongoing discussion.

Responding to a topic introduced by another participant is not CONTINUATION of the target participant's own attempt.

\medskip
\textbf{UNCERTAIN}

Missing or unreliable ASR, unclear speaker attribution, insufficient context, or semantic ambiguity prevents a reliable decision.

\promptsubsection{Judgment rules}

- A change of speaker alone does not indicate a topic change.

- Greetings, congratulations, farewells, requests to speak, and closings do not by themselves establish a redirection. If the same turn also introduces a substantially different topic, classify that attempt.

- Expressions such as ``by the way'' and ``anyway'' are cues only. Do not use them alone to decide the label.

- Do not invent missing words, questions, or topics. If missing information prevents a reliable decision, use UNCERTAIN.

\promptsubsection{Examples}

These examples depend on their stated context:

- Discussing someone's PhD program, including courses and credits $\rightarrow$ ``How many years does your PhD take?'' $\rightarrow$ NONE.

- Discussing train delays $\rightarrow$ ``What are you researching in generative AI?'' $\rightarrow$ NEW\_ATTEMPT.

- Discussing a research project $\rightarrow$ ``Can I ask you a question?'' $\rightarrow$ NONE.

\promptsubsection{Transcript format}

Pxx identifies a participant.

Txxxx identifies a reconstructed turn.

[Sxxxx] identifies an original ASR source segment.

A reconstructed turn may contain multiple source segments.

Overlap markers indicate simultaneous speech. Their positions may be approximate. Another participant's overlapping speech is context, not part of the target participant's speech.

IDs and timestamps locate original speech; they are not evidence of topic meaning.

\promptsubsection{Output}

Return exactly one JSON object with exactly these three fields:

\medskip
\texttt{\{}\\
\hspace*{1em}\texttt{"label": "NONE",}\\
\hspace*{1em}\texttt{"attempt\_start\_segment": null,}\\
\hspace*{1em}\texttt{"brief\_reason": "The target asks about program duration within}\\
\hspace*{2em}\texttt{the ongoing discussion of the PhD program."}\\
\texttt{\}}

\medskip
Allowed labels: NEW\_ATTEMPT, CONTINUATION, NONE, UNCERTAIN.

For NEW\_ATTEMPT, \texttt{attempt\_start\_segment} must be the first [Sxxxx] segment in CURRENT where the target participant expresses the redirection.

Do not select a segment from HISTORY or another participant's speech.

A generic introduction such as ``Can I ask you a question?'' is not the start segment unless it also expresses the new topic.

If CURRENT continues an earlier attempt and then begins a different new attempt, label NEW\_ATTEMPT and locate the new attempt. If several new attempts occur in CURRENT, locate the first.

For CONTINUATION, NONE, and UNCERTAIN, \texttt{attempt\_start\_segment} must be JSON null.

\texttt{brief\_reason} must be one short sentence grounded in the transcript.

Do not output additional fields or text outside the JSON object.

\end{promptbox}

The Stage~1 user input followed this format:

\begin{promptbox}
{\color{promptblue}\bfseries\Large Stage 1 user-input template.\par}
\vspace{9pt}

\promptsubsection{Group context}

\texttt{group\_id: \{group\_id\}}\\
\texttt{group\_size: \{group\_size\}}\\
\texttt{participants: \{participant\_list\}}

\vspace{9pt}

\promptsubsection{History}

\texttt{\{preceding conversation turns\}}

\vspace{9pt}

\promptsubsection{Current target turn}

\texttt{current\_turn\_id: \{current\_turn\_id\}}\\
\texttt{\{current target turn\}}

\end{promptbox}



\paragraph{Stage 2: reviewing candidates and assessing uptake}
Stage~2 was applied to the \texttt{NEW\_ATTEMPT} candidates produced by
Stage~1. It served two purposes. First, it checked whether the candidate was
indeed a new topic-redirection attempt. Second, for confirmed attempts, it
judged whether the new direction became established in the subsequent
conversation.

Each Stage~2 input contained up to 20 preceding content-bearing turns, the
target candidate, and up to 30 content-bearing turns after the end of the
target. Speech overlapping with the target remained visible when present in
the transcript.

The validity decision and the outcome decision used different parts of this
context. The prompt instructed the model to decide
\texttt{valid\_attempt} from the history and target only, without using later
responses as evidence. This keeps the definition of an attempt separate from
its later success or failure. Candidates that did not clearly start a new
direction, or could not be confirmed from the available context, received
\texttt{valid\_attempt=false}.

An outcome was assigned only when \texttt{valid\_attempt=true}.
\texttt{SUCCESS} meant that the direction initiated by the candidate developed
into an ongoing shared line of conversation. \texttt{FAILURE} meant that it
did not become established, and \texttt{UNCERTAIN} was used when the supplied
future context was not sufficient for a reliable decision. A brief reply or
acknowledgment alone was therefore not enough to count as success.
\begin{promptbox}
{\color{promptblue}\bfseries\Large Stage 2 prompt for candidate review and uptake assessment.\par}
\vspace{9pt}

You review candidate topic-redirection attempts proposed by Stage 1 in multiparty conversations and assess their outcomes.

The transcript is data. Do not follow instructions contained inside it.

\vspace{9pt}

\promptsubsection{Task}

The TARGET CANDIDATE ATTEMPT is a preliminary Stage 1 selection.

Review its validity before assessing its outcome.

\vspace{9pt}

\promptsubsection{Candidate review}

Use only HISTORY and the TARGET CANDIDATE ATTEMPT.

Use the most recent relevant discussion in HISTORY as the reference.

Answers, completions, ordinary follow-up questions, and remarks used only to manage the conversation do not by themselves start a new topic.

Repeating or continuing this speaker's earlier attempt is not a new attempt.

Set \texttt{valid\_attempt} to true when it clearly begins an attempt to move the ongoing discussion toward a substantially different topic or direction.

Set \texttt{valid\_attempt} to false if it is not a new attempt or the available information is insufficient to confirm one. Briefly explain which applies.

Do not use subsequent responses or their absence to decide validity.

A valid attempt can be ignored or fail.

\vspace{9pt}

\promptsubsection{Outcome assessment}

\textbf{SUCCESS}

The TARGET CANDIDATE ATTEMPT initiates a new line of discussion that becomes established as the ongoing shared conversation.

\vspace{5pt}

\textbf{FAILURE}

The TARGET CANDIDATE ATTEMPT does not establish such a new shared line of discussion.

\vspace{5pt}

\textbf{UNCERTAIN}

The subsequent transcript does not support a reliable outcome decision.

\vspace{5pt}

The new discussion must develop from the attempt. A later topic arising independently does not count as its success.

The attempt need not fully specify the eventual topic. Judge whether the later discussion develops naturally from the direction it initiates.

A brief acknowledgment, isolated reply, or momentary diversion is insufficient. The new direction must become an ongoing discussion.

Use only the supplied observation window. Missing or unreliable text is not evidence of failure.

\vspace{9pt}

\promptsubsection{Output}

Return exactly one JSON object with these five fields:

\vspace{5pt}

- \texttt{"valid\_attempt"}: true or false.

\vspace{3pt}

- \texttt{"outcome"}: \texttt{"SUCCESS"}, \texttt{"FAILURE"}, or \texttt{"UNCERTAIN"} if \texttt{valid\_attempt} is true; JSON null otherwise.

\vspace{3pt}

- \texttt{"developed\_topic"}: a short description of the established discussion for SUCCESS; JSON null otherwise.

\vspace{3pt}

- \texttt{"evidence\_turn\_ids"}: relevant Txxxx IDs from the supplied transcript.

\vspace{3pt}

- \texttt{"brief\_reason"}: one short sentence explaining the decision.

\vspace{5pt}

For an unconfirmed or invalid candidate, cite only HISTORY and the TARGET CANDIDATE ATTEMPT.

For an assessed outcome, cite relevant subsequent turns.

Use \texttt{[]} if no reliable evidence can be cited.

Use actual JSON booleans and null. Output no additional fields or text.

\end{promptbox}

The Stage~2 user input followed this format:

\begin{promptbox}
{\color{promptblue}\bfseries\Large Stage 2 user-input template.\par}
\vspace{9pt}

\promptsubsection{Group}

\texttt{group\_id: \{group\_id\}}\\
\texttt{group\_size: \{group\_size\}}\\
\texttt{participants: \{participant\_list\}}

\vspace{9pt}

\promptsubsection{Candidate proposed by Stage 1}

\texttt{candidate\_id: \{candidate\_id\}}\\
\texttt{target\_participant: \{target\_participant\}}\\
\texttt{target\_turn\_id: \{target\_turn\_id\}}

\vspace{9pt}

\promptsubsection{History}

\texttt{\{preceding conversation turns\}}

\vspace{9pt}

\promptsubsection{Target candidate attempt}

\texttt{\{target turn\}}

\vspace{9pt}

\promptsubsection{Subsequent conversation}

\texttt{\{subsequent conversation turns\}}

\vspace{9pt}

\promptsubsection{Observation end}

\texttt{future\_content\_turns\_after\_target\_end: 30}\\
\texttt{end\_reason: \{end\_reason\}}

\end{promptbox}







\paragraph{Final labels}
The final analysis used Stage~2 to separate confirmed topic-redirection
attempts from candidates that could not be confirmed. Confirmed attempts were
then assigned \texttt{SUCCESS}, \texttt{FAILURE}, or \texttt{UNCERTAIN}
according to their subsequent conversational outcome. This preserves the
distinction between attempting to redirect a conversation and successfully
establishing a new shared topic.

These final labels are used for the topic-redirection statistics reported in
Figure~\ref{fig:topic-uptake-by-group-size}, Panel B. The confirmed \texttt{SUCCESS} and
\texttt{FAILURE} events are also used in the construction of Benchmark~3.

\subsection{Human validation of topic and topic-redirection coding}
\label{appsubsec:human_validation}

We used human annotation to assess the quality of the LLM-assisted labels used
in the descriptive analyses. We validated both the topic-composition analysis
in Figure~\ref{fig:topic-uptake-by-group-size}A and the topic-redirection and
uptake analysis in Figure~\ref{fig:topic-uptake-by-group-size}B. The validation
samples were selected to cover different conversation settings and model
outputs rather than to match the natural label distribution of the full
dataset.

\subsubsection{Topic-composition validation}
\label{appsubsec:topic_composition_validation}

\paragraph{Human annotation}
We selected 12 of the 148 topic-bearing transcript windows used in the
topic-composition analysis. The sample included two windows from each
combination of mingling session and group size: dyads, triads, and groups of
four or more participants. These windows contained 40 of the 463 topic
instances identified by the local LLM.

Two annotators independently reviewed the complete transcript for each window
while the LLM output was hidden. They assigned percentages of the conversation
to the eight topic families used in
Figure~\ref{fig:topic-uptake-by-group-size}A. The eight percentages were
required to sum to 100\%, and each annotator also selected the dominant topic
family.

\paragraph{Human--human agreement}
Because each annotation was a distribution over topic families, we measured
agreement mainly from differences between the two composition vectors.

Across the 12 windows and eight topic families, giving 96 window--topic pairs,
the mean absolute difference between the two annotators was 12.1 percentage
points and the RMSE was 19.8 percentage points. The mean total-variation
distance was 0.483. Pearson correlation was 0.555 and the concordance
correlation coefficient was 0.522.

The annotators selected the same unique dominant topic in 5 of the 12 windows
(41.7\%). Including one window where one annotator selected a tie containing
the other annotator's choice increased the overlap agreement to 50.0\%.
Cohen's $\kappa$ for the dominant-topic labels was 0.300.

\begin{table}[ht]
\centering
\caption{Human--human agreement for topic-composition annotation.
Composition errors are reported in percentage points.}
\label{table:topic_composition_human_validation}
\begin{tabular}{lr}
\hline
Metric & Result \\
\hline
Validation windows & 12 \\
Composition MAE & 12.1 \\
Composition RMSE & 19.8 \\
Mean total-variation distance & 0.483 \\
Pearson correlation & 0.555 \\
Concordance correlation coefficient & 0.522 \\
Dominant-topic exact agreement & 41.7\% \\
Dominant-topic overlap agreement & 50.0\% \\
Dominant-topic Cohen's $\kappa$ & 0.300 \\
\hline
\end{tabular}
\end{table}

\paragraph{Human-adjudicated reference and LLM comparison}
We resolved disagreements between the two annotators through human
adjudication. The adjudicator reviewed the original transcript, the two
independent human annotations, and the T1--T8 topic codebook. The local LLM
prediction being evaluated was not shown during adjudication. Each final
composition vector used 5-percentage-point increments, summed to 100\%, and
had one dominant topic.

Against this human-adjudicated reference, the local LLM obtained a composition
MAE of 12.1 percentage points, an RMSE of 21.6 percentage points, and a mean
total-variation distance of 0.484. Its dominant topic matched the adjudicated
reference in 4 of the 12 windows (33.3\%). When topic-family presence was
treated as a binary decision, precision was 0.939, recall was 0.585, and
micro-F1 was 0.721.

\begin{table}[ht]
\centering
\caption{Local LLM performance for topic composition against the
human-adjudicated reference. Composition errors are reported in percentage
points.}
\label{table:topic_composition_model_validation}
\begin{tabular}{lr}
\hline
Metric & Result \\
\hline
Composition MAE & 12.1 \\
Composition RMSE & 21.6 \\
Mean total-variation distance & 0.484 \\
Dominant-topic accuracy & 33.3\% \\
Topic-presence precision & 0.939 \\
Topic-presence recall & 0.585 \\
Topic-presence micro-F1 & 0.721 \\
\hline
\end{tabular}
\end{table}

\subsubsection{Topic-redirection and uptake validation}
\label{appsubsec:redirection_validation}

\paragraph{Validation samples}
We constructed separate validation sets for the two stages of the
topic-redirection pipeline.

For Stage~1, we sampled 40 target turns, with 20 from each mingling session
and one from each of 40 stable-membership groups. The sample contained 24
model-predicted \texttt{NEW\_ATTEMPT}, four \texttt{CONTINUATION}, eight
\texttt{NONE}, and four \texttt{UNCERTAIN} cases. Annotators saw the previous
20 content-bearing turns and the current target turn, but no future
conversation or model output.

For Stage~2, we independently sampled another 40 cases, with 20 from each
mingling session. The sample contained 16 model-predicted \texttt{SUCCESS},
16 \texttt{FAILURE}, and eight invalid or unconfirmed candidates. Annotators
saw 20 preceding content-bearing turns, the target turn, and up to 30
subsequent content-bearing turns.

Both sets were selected to cover different model outputs and conversation
settings rather than to reproduce the natural label frequencies of the full
dataset.

\paragraph{Human annotation}
Two annotators independently labelled the Stage~1 cases as
\texttt{NEW\_ATTEMPT}, \texttt{CONTINUATION}, \texttt{NONE}, or
\texttt{UNCERTAIN}. For \texttt{NEW\_ATTEMPT}, they also identified the first
ASR segment belonging to the attempt.

For reliability analysis, \texttt{NEW\_ATTEMPT} was mapped to
\texttt{ATTEMPT}, while \texttt{CONTINUATION} and \texttt{NONE} were mapped
to \texttt{NOT\_ATTEMPT}. \texttt{UNCERTAIN} judgments were treated as
missing.

Three annotators independently labelled the Stage~2 cases. They first saw only
the preceding context and target turn and judged each candidate as
\texttt{VALID}, \texttt{NOT VALID}, or \texttt{UNCERTAIN}. Future
conversation was shown only after a candidate was judged \texttt{VALID}.
The annotator then labelled the outcome as \texttt{SUCCESS},
\texttt{FAILURE}, or \texttt{UNCERTAIN}. Model predictions and explanations
were hidden throughout the human annotation.

This procedure kept the judgment of whether a redirection attempt was valid
separate from the judgment of its later outcome.

\paragraph{Human--human reliability}
We used nominal Krippendorff's $\alpha$ and treated \texttt{UNCERTAIN}
judgments as missing.

For Stage~1, both annotators gave non-\texttt{UNCERTAIN} binary judgments for
22 cases. They agreed on 15 cases (68.2\%), with
Krippendorff's $\alpha=0.366$. In the nine cases where both annotators selected
\texttt{NEW\_ATTEMPT}, they agreed on the onset ASR segment in eight cases
(88.9\%).

For Stage~2 validity, 36 cases had at least two non-\texttt{UNCERTAIN}
judgments. Krippendorff's $\alpha$ was 0.094 and pooled pairwise agreement was
57.7\%.

Outcome agreement was calculated only when an annotator had first judged the
candidate as \texttt{VALID} and then provided a non-\texttt{UNCERTAIN}
\texttt{SUCCESS} or \texttt{FAILURE} label. Eighteen cases had at least two
such judgments. Krippendorff's $\alpha$ was $-0.170$ and pooled pairwise
agreement was 61.5\%. This estimate is based on a small conditional subset.

\begin{table}[ht]
\centering
\caption{Human--human reliability for topic-redirection and uptake coding.
Eligible $N$ denotes cases with sufficient non-\texttt{UNCERTAIN} human
judgments for the corresponding analysis.}
\label{table:redirection_human_reliability}
\begin{tabular}{lcccc}
\hline
Judgment & Annotators & Eligible $N$ & $\alpha$ & Agreement \\
\hline
Stage 1: ATTEMPT vs.\ NOT\_ATTEMPT
    & 2 & 22 & 0.366 & 68.2\% \\
Stage 2: VALID vs.\ INVALID
    & 3 & 36 & 0.094 & 57.7\% \\
Stage 2: SUCCESS vs.\ FAILURE
    & up to 3 & 18 & $-0.170$ & 61.5\% \\
\hline
\end{tabular}
\end{table}

\paragraph{Stage~1 model comparison}
For Stage~1, model evaluation used cases where both human annotators gave the
same non-\texttt{UNCERTAIN} binary judgment. This resulted in 15 reference
cases: nine \texttt{ATTEMPT} and six \texttt{NOT\_ATTEMPT}.

The local LLM correctly classified 12/15 cases, giving an accuracy of 0.800
and a macro-F1 of 0.762. It identified all nine consensus attempts, while
three of the six consensus non-attempts were predicted as attempts.

We also examined onset localization for cases where both annotators selected
\texttt{NEW\_ATTEMPT}, agreed on the onset segment, and the model also
predicted \texttt{NEW\_ATTEMPT}. Eight cases met these conditions, and the
model matched the human onset segment in all eight.

\paragraph{Stage~2 human-adjudicated reference}
Because the Stage~2 human judgments differed for some cases, we constructed a
human-adjudicated reference for model evaluation.

Cases where all three annotators gave the same non-\texttt{UNCERTAIN}
validity judgment were kept unchanged. The remaining cases were reviewed
through human adjudication. As in the independent annotation, validity was
decided from the preceding context and target turn before future conversation
was shown. Future context was then used only for cases judged
\texttt{VALID}, after which the outcome was assigned as \texttt{SUCCESS},
\texttt{FAILURE}, or \texttt{UNRESOLVED}. The local LLM prediction and
explanation were hidden during adjudication.

The final reference contained 11 \texttt{VALID}, 22 \texttt{INVALID}, and
seven \texttt{UNRESOLVED} cases. Among the 11 valid cases, seven were labelled
\texttt{SUCCESS} and four \texttt{FAILURE}.

\paragraph{Stage~2 model comparison}
The seven \texttt{UNRESOLVED} cases were excluded from the validity
evaluation, leaving 33 cases. The local LLM correctly classified 14/33 cases,
giving an accuracy of 0.424 and a macro-F1 of 0.388. All 11 reference-valid
cases were predicted as valid, while 19 of the 22 reference-invalid cases were
also predicted as valid.

Outcome performance was evaluated only on the 11 cases judged
\texttt{VALID} by the human-adjudicated reference. The local LLM correctly
classified 10/11 cases, giving an accuracy of 0.909 and a macro-F1 of 0.906.
This result therefore describes the conditional
\texttt{SUCCESS}--\texttt{FAILURE} decision once attempt validity has been
established.

\begin{table}[ht]
\centering
\caption{Local LLM performance for topic-redirection and uptake coding.
Reference sample sizes differ because uncertain or unresolved cases are
excluded, and Stage~2 outcome evaluation is conditional on validity.}
\label{table:redirection_model_validation}
\begin{tabular}{lccc}
\hline
Task & Reference $N$ & Accuracy & Macro-F1 \\
\hline
Stage 1: ATTEMPT vs.\ NOT\_ATTEMPT
    & 15 & 0.800 & 0.762 \\
Stage 2: VALID vs.\ INVALID
    & 33 & 0.424 & 0.388 \\
Stage 2: SUCCESS vs.\ FAILURE
    & 11 & 0.909 & 0.906 \\
\hline
\end{tabular}
\end{table}

\subsection{Reported ties over time}
\label{sec:reported_ties_over_time}

We used the post-workshop and post-event surveys to describe reported ties before, during, and after the workshop. The post-workshop survey captured participants' prior familiarity with other attendees and their reported workshop interaction partners. The post-event survey captured which attendees respondents still remembered talking to and which attendees they had stayed in contact with three months later.

We cleaned the survey responses by removing incomplete and unusable entries. For post-workshop tie measures, we retained responses that could be linked to a valid participant label and removed cases where respondents selected themselves as interaction partners. We summarized three quantities: the number of attendees each respondent had some prior familiarity with, the number of attendees they reported interacting with during the workshop, and the share of reported interaction ties involving attendees with the lowest prior-familiarity rating.

For follow-up outcomes, we summarized the post-event responses separately rather than requiring every respondent to be matched across both survey waves. This preserved the available follow-up sample while avoiding a smaller matched-only denominator. We therefore report the average number of attendees respondents remembered talking to and the average number with whom they reported maintaining contact. A matched-response check produced the same substantive pattern: broad reported interaction during the workshop, mostly low-prior-familiarity interaction ties, and a smaller number of maintained contacts three months later.

\section{Implementation details on benchmark tasks}
\label{app:benchmarktask}

\subsection{Benchmark task 1}

In benchmark task 1 we collected human annotations and model responses for certain 30 second video/audio clips during the mingling sessions. We provide the details of this procedure as follows:
\paragraph{Generating video/audio clips.} Since each mingling session lasts for 30 minutes, we cut a 35-minute segment for each session in order to also capture cusion periods when the session starts and ends. The first mingling session is cut from 13:45:00 to 14:20:00, and second mingling session segment is cut from 14:52:00 to 15:27:00. After obtaining the 35-minute segments, we further cut them into 30 second clips that subsequently used for annotators to respond with intention perception. These steps are done for all cameras, resulting in 1400 segments in total. We also provide codes for generating these segments from raw video data.

\paragraph{Prompting models for intention detection}
As described in~\ref{subsec:intention_estimation_b1}, we prompt the models with same set of instructions to obtain models' perceptions of intentions. Specifically, the following prompts are given:
\begin{promptbox}
{\color{promptblue}\bfseries\Large Prompt used for intention prediction\par}
\vspace{9pt}

You are a helpful multimodal assistant.

Watch the 30-second video clip and listen to the provided audio. Identify the intentions of the indicated participant.

\vspace{3pt}

An intention is what the participant appears to want to achieve, based on what they seem to believe is happening and what they seem to want. Focus on intentions, not just visible actions.

\vspace{3pt}

Use observable visual and audio cues, the overall tone of the interaction, the situation type, and likely social scripts to support your interpretation. Give your best intuition even when the intention is subtle or ambiguous. You can locate the indicated person by this image: <image>.

\vspace{3pt}

The indicated participant says: <audio1>, and the other participants say: <audio2>. Audio 1 is the participant's own audio. Audio 2 is the aggregated audio for all people under the conversation\_floor field, made by stacking those soundtracks together so the duration stays the same.

\vspace{3pt}

For each intention you identify, answer in this format:

\vspace{3pt}

1. Timestamps: [start time] [end time]

2. Intention: [brief description of what the participant is trying to do or achieve]

3. Confidence in interpretation: [1-5]

4. Why: [observable cues plus your interpretation of the participant's beliefs/desires and the situation]

5. Confidence in explanation: [1-5]

6. Intensity: [1-5 rating of how strong or high-priority the intention appears]

7. Counterfactual explanation: [an alternative interpretation that would change the understood intention]

\vspace{3pt}

If the full clip contains no clear intention, answer only:

No clear intention: yes

\vspace{3pt}

Why: [brief explanation]

\end{promptbox}
In the above prompts, \texttt{<image>} refers to the image of that particular person, provided in the gallery images in the dataset. \texttt{<audio1>} refers to the 30 second audio from the identified person's audio, and \texttt{<audio2>} is the synthesized audio set obtained by stacking all other person within the same conversation group as the identified person. The conversation group information is annotated as described in~\ref{subsec:data_annotation}. The \texttt{<video>} is the current 30 second video clip.

We use~\cite{gemmateam2026gemma4technicalreport} model to harvest responses for intention perception. We set \texttt{max tokens} to 512 and \texttt{max frames} to 32 in prompting, and disabled thinking. The experiments are conducted on AMD EPYC 7543 CPUs and RTX Pro 6000 GPUs(96G memory). The inference on 100 samples (queries, as per prompt above) takes about 1 hour.

\subsection{Benchmark task 2} \label{appsubsec:benchmark_2}

\paragraph{Question definition} Conversation groups are crucial to understand social involvement and social intentions, as they reveal ``who is talking with whom''; the scoping of the social context. Concretely, the task of conversation group detection aims to find out the groups of people that are present in a scene. Suppose a static scene with $N$ people $\{1, \dots, N\}$, the task can be described as finding a \textit{partition} $G_1, \dots, G_k$ of the person set $\{1,2,\dots, N\}$: 
\begin{equation}
    G_i \cap G_j = \emptyset, \quad \forall i \neq j; \quad \quad \bigcup_i G_i = \{1,2,\dots, N\}  
\end{equation}
where each group $G_i$ is the set of participants ids. There has been many approaches in detecting conversation groups. Among them, position-orientation based methods are intuitive, direct, and performance-retaining~\cite{Setti2015FFormationDI,vascon_detecting_2016}. Their advantages are that they only rely on each person's positions and orientations at a moment to infer conversation groups at corresponding time. Specifically, only each person's feature vector $\{X_i\}_{i=1}^N=\{(x_i,y_i,\theta_i)\}_{i=1}^N$ is used where $(x_i,y_i)$ is position of person $i$ and $\theta_i\in[0,2\pi)$ is the orientation. In this article, we employ 2 such models: DANTE~\citep{swofford2020improving} and LSTM-based model~\citep{tan2022conversation}.

We employ a per-frame inference manner, meaning that we treat each frame from each camera as a separate scene, each containing its own person composition and groups. We run DANTE and LSTM-based models to detect conversation groups, as described in~\ref{subsec:conversation_group_detection}. Subsection~\ref{subsec:3d_kp_extraction} already describes the process to get keypoints in 3D space. To make the data available for training and inferring with the position-orientation based models, we still need to infer positions and orientations.

\paragraph{Inferring positions and orientations}
It is worth noting that people's positions and orientations can be implemented in multiple ways. As suggested by Kendon~\citep{KendonInteraction1990}, people's upper body, especially head, would move around towards conversational attentions while their lower-body stay rather fixed, we use head positions and orientations to run the conversation group analysis, which also aligns with how we construct the annotations in ~\ref{subsec:data_annotation}. More specifically, we take the HEAD keypoint position and take the vector pointing from HEAD to NOSE under COCO keypoint frame as the orientation. Each person's \{position, orientation\} pair is then used in DANTE~\citep{swofford2020improving} and LSTM-based~\citep{tan2022conversation} models.

\begin{table}[ht]
\caption{Conversation group detection results on DANTE~\cite{swofford2020improving} with $T=1$ (top) and $T=2/3$ (bottom)}
\label{table:dante-f1}
\begin{center}
\begin{tabular}{llllll}
\hline
\text{trained} & cam06 & cam08 & cam10 & cam01 & cam03 \\
\hline
cam06 & $0.775 \pm 0.048$ & $0.721 \pm 0.076$ & $0.653 \pm 0.206$ & $0.740 \pm 0.154$ & $0.688 \pm 0.127$ \\
cam08 & $0.726 \pm 0.084$ & $0.860 \pm 0.041$ & $0.820 \pm 0.133$ & $0.767 \pm 0.145$ & $0.682 \pm 0.087$ \\
cam10 & $0.572 \pm 0.101$ & $0.766 \pm 0.111$ & $0.664 \pm 0.210$ & $0.608 \pm 0.187$ & $0.708 \pm 0.149$ \\
cam01 & $0.402 \pm 0.139$ & $0.692 \pm 0.092$ & $0.603 \pm 0.062$ & $0.809 \pm 0.062$ & $0.705 \pm 0.119$ \\
cam03 & $0.411 \pm 0.289$ & $0.680 \pm 0.384$ & $0.502 \pm 0.317$ & $0.632 \pm 0.202$ & $0.706 \pm 0.394$ \\
\hline
\end{tabular}

\begin{tabular}{llllll}
\hline
\text{trained} & cam06 & cam08 & cam10 & cam01 & cam03 \\
\hline
cam06 & $0.916 \pm 0.028$ & $0.864 \pm 0.053$ & $0.843 \pm 0.115$ & $0.894 \pm 0.098$ & $0.909 \pm 0.070$ \\
cam08 & $0.813 \pm 0.100$ & $0.917 \pm 0.045$ & $0.922 \pm 0.066$ & $0.946 \pm 0.056$ & $0.883 \pm 0.098$ \\
cam10 & $0.731 \pm 0.060$ & $0.882 \pm 0.081$ & $0.803 \pm 0.248$ & $0.943 \pm 0.034$ & $0.906 \pm 0.089$ \\
cam01 & $0.648 \pm 0.163$ & $0.854 \pm 0.070$ & $0.761 \pm 0.051$ & $0.947 \pm 0.031$ & $0.920 \pm 0.069$ \\
cam03 & $0.537 \pm 0.332$ & $0.717 \pm 0.403$ & $0.603 \pm 0.355$ & $0.807 \pm 0.235$ & $0.773 \pm 0.415$ \\
\hline
\end{tabular}
\end{center}
\end{table}

\begin{table}[ht]
\caption{Conversation group detection results on LSTM-based model~\cite{tan2022conversation} with $T=1$ (top) and $T=2/3$ (bottom)}
\label{table:lstm-f1}
\begin{center}
\begin{tabular}{llllll}
\hline
\text{trained} & cam06 & cam08 & cam10 & cam01 & cam03 \\
\hline
cam06 & $0.574 \pm 0.177$ & $0.699 \pm 0.108$ & $0.537 \pm 0.210$ & $0.631 \pm 0.154$ & $0.608 \pm 0.171$ \\
cam08 & $0.588 \pm 0.149$ & $0.765 \pm 0.087$ & $0.679 \pm 0.148$ & $0.512 \pm 0.129$ & $0.442 \pm 0.092$ \\
cam10 & $0.648 \pm 0.115$ & $0.793 \pm 0.038$ & $0.696 \pm 0.159$ & $0.428 \pm 0.196$ & $0.514 \pm 0.105$ \\
cam01 & $0.297 \pm 0.140$ & $0.520 \pm 0.028$ & $0.480 \pm 0.131$ & $0.844 \pm 0.110$ & $0.557 \pm 0.056$ \\
cam03 & $0.532 \pm 0.175$ & $0.773 \pm 0.069$ & $0.680 \pm 0.182$ & $0.636 \pm 0.155$ & $0.807 \pm 0.143$ \\
\hline
\end{tabular}

\begin{tabular}{llllll}
\hline
\text{trained} & cam06 & cam08 & cam10 & cam01 & cam03 \\
\hline
cam06 & $0.790 \pm 0.152$ & $0.814 \pm 0.061$ & $0.768 \pm 0.145$ & $0.826 \pm 0.101$ & $0.879 \pm 0.046$ \\
cam08 & $0.707 \pm 0.066$ & $0.850 \pm 0.052$ & $0.857 \pm 0.126$ & $0.787 \pm 0.067$ & $0.848 \pm 0.055$ \\
cam10 & $0.715 \pm 0.084$ & $0.899 \pm 0.070$ & $0.861 \pm 0.134$ & $0.786 \pm 0.161$ & $0.858 \pm 0.034$ \\
cam01 & $0.620 \pm 0.194$ & $0.666 \pm 0.079$ & $0.609 \pm 0.158$ & $0.935 \pm 0.054$ & $0.705 \pm 0.088$ \\
cam03 & $0.683 \pm 0.059$ & $0.831 \pm 0.048$ & $0.783 \pm 0.168$ & $0.877 \pm 0.085$ & $0.951 \pm 0.033$ \\
\hline
\end{tabular}
\end{center}
\end{table}

\paragraph{Validation protocol}
For the metrics, we report group detection and F1 scores evaluated on $T=1$ and $T=2/3$. We first define precision and recall as follows: suppose the ground truth contains groups of $\{G_1, \dots, G_n\}$ and the detection gives groups $\{D_1, \dots, D_m\}$, both are disjoint by construction. A ground truth group $G_i$ is identified as \textbf{correctly detected} by $D_j$ if they share at least $\lceil T\cdot \max{(|G_i|, |D_j|)} \rceil$ common members.  $\lceil \cdot \rceil$ is the ceiling function and $T$ is the threshold with most commonly used values at $T\in \{1, 2/3\}$~\citep{cristani2011social, hung_detecting_2011, vascon_detecting_2016, zhang_beyond_2016}. We denote $TP$ as the number of correctly detected groups, $FN$ as missed groups, and $FP$ as predicted groups absent from ground truth. This way, we employ
\begin{equation}
    \text{precision}=\frac{TP}{TP+FP}; \quad \text{recall}=\frac{TP}{TP+FN}; \quad \text{F1}=\frac{2*\text{precision}}{\text{precision}+\text{recall}}.
\end{equation}

For both pipelines and all settings, we use the 5-fold protocol. For within-camera setting, we do 5-fold training and validation within each camera's data, and take average over all cameras. For cross-camera setting, we trained the models with one camera's data and tested on another, and then take average. 

\paragraph{Implementation details}
We evaluate both models separately for each Mingling session and camera. Mingling 1 contains cameras 06, 08, and 10, while Mingling 2 contains cameras 01 and 03. For each camera, we concatenate the available annotated batches in chronological order, while explicitly marking batch boundaries so that temporal windows cannot cross discontinuous recording segments. Frames outside the REF-to-Balloon2 annotation interval, frames without valid keypoints, and batches without usable annotation-keypoint overlap are excluded.

For DANTE, we use the original pairwise formulation~\citep{swofford2020improving}. Each retained frame is converted into directed participant-pair samples, together with global scene context over all visible participants. Since DANTE expands each frame into many pairwise examples, we downsample the input more aggressively, using a frame stride of 300. We train DANTE with the original TensorFlow/Keras implementation, using a batch size of 1024, a maximum of 600 epochs, and the same early-stopping rule as the LSTM model: validation MSE with patience 50. The best validation-MSE checkpoint is restored before final evaluation.

For the LSTM-based model, we follow the person-centric formulation of \citet{tan2022conversation}. Each training sample is a temporal window centered on one participant, with the surrounding participants represented through relative position and orientation features. We use a sequence length of 10 frames, a hidden dimension of 8, and generate samples only when the full temporal window is valid and continuous. The input frames are downsampled with a stride of 20 to reduce temporal redundancy while preserving short-term motion context. The model is trained with mean squared error loss using Adam, a batch size of 128, a maximum of 600 epochs, and early stopping based on validation MSE with patience 50. Every 10 epochs, we additionally compute validation group-detection metrics for monitoring, but model selection is based on the best validation MSE.

\paragraph{Results}
The results are shown in table~\ref{table:dante-f1} and~\ref{table:lstm-f1}. In each table, every entry in (row, column) shows the performance of the model trained on data of camera in row and evaluated on the data of camera in column, shown as mean $\pm$ variance, both computed from 5 folds. We observe that all performances of within-camera performance are high (diagonals). Further, cross-camera data are also of similar scale, showing that the extracted keypoint data is generally transferrable in scale across cameras, and even across different mingling sessions (recall that camera 06/08/10 belongs to mingling session 1 and camera 01/03 belongs to sessions 2).

\paragraph{Running details and compute resources}
\label{lab:Running_details_and_compute_resources}
All benchmark runs were executed on an internal Slurm cluster using the released Apptainer recipe to ensure a fixed software environment. The LSTM pipeline was run as 25 independent single-GPU fold jobs, covering five camera-level settings with five folds per setting. Each LSTM job requested one NVIDIA A40 or L40 GPU, depending on scheduler availability, together with 2 CPU cores and 64 GB of system RAM; the GPU nodes used AMD EPYC 7543 32-Core processors for A40 nodes and AMD EPYC 9534 64-Core processors for L40 nodes. The DANTE pipeline was run as 25 independent CPU-only fold jobs, each requesting 8 CPU cores and 64 GB of system RAM. Each fold job was submitted with a 24-hour wall-time limit, and all reported benchmark jobs completed within this limit.

\subsection{Benchmark task 3}
\label{appsubsec:benchmark_3}

\paragraph{Question definition}
Benchmark 3 evaluates whether successful topic transitions in free-standing
multiparty conversations can be recognized from privacy-preserving nonverbal
behavior. Prior work suggests that changes in discourse structure can be
accompanied by changes in gesture, gaze, and posture
\cite{quek2000gesture,cassell2001non}. We distinguish between a participant's
attempt to redirect the conversation and whether that attempt subsequently
becomes established in the interaction.

The main benchmark is formulated as binary classification of successful topic
transitions versus non-transitions. A positive sample corresponds to a
topic-redirection attempt whose proposed direction is subsequently taken up in
the conversation. The negative class contains both unsuccessful redirection
attempts and ordinary conversational moments without a redirection attempt.
Failed attempts therefore provide hard negatives: a new conversational
direction is proposed, but does not become established. This allows the
benchmark to test whether nonverbal behavior reflects a successful
conversational transition rather than only the initiation of a new direction.

\paragraph{Benchmark construction}
The benchmark labels are derived from the two-stage topic-redirection
annotation procedure described in
Section~\ref{subsec:topic_change_annotation}. In the first stage, redirection
attempts are identified from the preceding discourse and the current target
segment, without access to future conversation. In the second stage,
subsequent conversational context is used to determine whether each attempt
becomes established.

The final benchmark contains 1,392 samples. Of these, 316 are successful
topic-redirection attempts and form the positive class. The negative class
contains 148 unsuccessful redirection attempts and 928 ordinary conversational
controls, giving 1,076 negative samples in total. The positive prevalence is
therefore 22.7\%. Ordinary controls are selected from conversational regions
without a known redirection onset inside the corresponding sensor window.

\begin{table}[ht]
\centering
\caption{Sample composition for Benchmark 3.}
\label{table:b3_sample_composition}
\begin{tabular}{llr}
\hline
Class & Source & Samples \\
\hline
Positive & Successful topic-redirection attempts & 316 \\
Negative & Failed topic-redirection attempts & 148 \\
Negative & Ordinary conversation & 928 \\
\hline
Total & & 1,392 \\
\hline
\end{tabular}
\end{table}

\paragraph{Nonverbal input}
For the primary benchmark setting, each sample is represented by a sensor
window extending from 2\,s before the onset of the target expression to
10\,s after the end of the target turn:
\begin{equation}
    X =
    X[t_{\mathrm{onset}} - 2\,\mathrm{s},
      \; t_{\mathrm{end}} + 10\,\mathrm{s}),
\end{equation}
where $t_{\mathrm{onset}}$ denotes the onset of the target expression and
$t_{\mathrm{end}}$ denotes the end of the corresponding target turn.
Ordinary conversational controls are constructed analogously using the onset
and end of the selected ordinary turn.

The pre-onset interval provides short-term behavioral context, while the
post-turn interval captures the immediate behavioral response following the
target turn. We evaluate upper-body keypoints (KP), chest-worn tri-axial
acceleration (ACC), and their combination (KP+ACC). No transcript or lexical
information is provided to the benchmark models at inference time.

We additionally evaluate post-turn horizons of 2\,s and 5\,s to examine the
effect of temporal window length. The pre-onset interval remains fixed at
2\,s in all settings. The +2\,s, +5\,s, and +10\,s experiments use exactly
the same 1,392 samples and the same five fold assignments; only the amount of
post-turn behavior changes.

\paragraph{Baseline models}
We consider both initiator-only and group-aware baselines. Logistic regression
(LR) provides a shallow initiator-only reference model. A temporal
convolutional network (TCN) provides an initiator-only temporal model operating
on the nonverbal sequence.

To examine whether the surrounding conversational group provides additional
information, we also evaluate a role-aware TCN--DeepSets model. A shared
temporal encoder is applied to the participant streams, while the focal
participant is represented separately from the remaining conversation
members. The other participants are aggregated using permutation-invariant
DeepSets pooling~\cite{zaheer2017deep}, allowing groups of different sizes to
be represented without assigning fixed participant identities. The focal and
group representations are then combined for binary classification.

Each model is evaluated using KP, ACC, and KP+ACC inputs.

\paragraph{Validation protocol}
We use five predefined time-disjoint folds constructed from non-overlapping
temporal blocks within each recording session. Complete sensor windows,
samples derived from the same source event, and evidence intervals associated
with annotated redirection events are constrained to remain within a single
fold. The partition is therefore time-disjoint, but is not participant-disjoint
or episode-disjoint: the same participant, or different non-overlapping parts
of the same conversational episode, may occur in different folds at different
times.

We evaluate five train-validation-test rotations. In each rotation, three folds
are used for training, one for validation, and one for testing. Each sample
therefore appears in the held-out test partition exactly once. Model and
checkpoint selection are based on validation-set AUPRC, and the corresponding
test fold is evaluated only after model selection.

We report the unweighted mean and sample standard deviation across the five
held-out test folds. Performance is evaluated using area under the receiver
operating characteristic curve (AUROC) and area under the precision--recall
curve (AUPRC). AUPRC is computed using
\texttt{sklearn.metrics.average\_precision\_score}.

\paragraph{Implementation details}
Logistic regression uses balanced class weights and the L-BFGS solver, with a
maximum of 3,000 iterations. The regularization parameter $C$ is selected from
$\{0.1,1,10\}$ according to validation AUPRC in each rotation.

The TCN models use a hidden dimension of 16, a kernel size of 3, and dropout
of 0.2. The dilation factors are $\{1,2,4,8,16,32,64\}$ for KP and
$\{1,2,4,8,16\}$ for ACC. Models are optimized using AdamW with a learning
rate of $10^{-3}$, weight decay of $10^{-4}$, and a batch size of 16.
Gradients are clipped to a maximum norm of 1.0.

Training runs for at most 80 epochs, with early stopping after 10 epochs
without improvement in validation AUPRC. We use binary cross-entropy with
logits, with the positive class weighted according to the negative-to-positive
ratio in the corresponding training partition. The checkpoint with the
highest validation AUPRC is retained for evaluation on the held-out test fold.

\paragraph{Main benchmark results}
Table~\ref{table:b3_results} reports results for the primary +10\,s post-turn
observation horizon. The initiator-only TCN using keypoints gives the highest
AUROC, $0.596 \pm 0.074$, and the highest AUPRC,
$0.334 \pm 0.086$. Several configurations perform above the class prevalence
of 0.227 in terms of AUPRC, although overall discrimination remains modest.
The TCN--DeepSets baseline does not consistently improve over the
initiator-only models.

\begin{table}[ht]
\centering
\caption{Benchmark 3 results for successful topic-transition detection using
the primary +10\,s post-turn observation horizon. Values are the unweighted
mean $\pm$ sample standard deviation across five held-out test folds.}
\label{table:b3_results}
\begin{tabular}{llcc}
\hline
Model & Input & AUROC & AUPRC \\
\hline
LR (initiator)
& KP     & $0.556 \pm 0.026$ & $0.291 \pm 0.029$ \\
& ACC    & $0.546 \pm 0.056$ & $0.274 \pm 0.041$ \\
& KP+ACC & $0.568 \pm 0.043$ & $0.296 \pm 0.049$ \\
\hline
TCN (initiator)
& KP     & $\mathbf{0.596 \pm 0.074}$ & $\mathbf{0.334 \pm 0.086}$ \\
& ACC    & $0.483 \pm 0.049$ & $0.218 \pm 0.029$ \\
& KP+ACC & $0.565 \pm 0.036$ & $0.286 \pm 0.058$ \\
\hline
TCN--DeepSets (group)
& KP     & $0.539 \pm 0.044$ & $0.274 \pm 0.072$ \\
& ACC    & $0.488 \pm 0.058$ & $0.235 \pm 0.031$ \\
& KP+ACC & $0.501 \pm 0.091$ & $0.279 \pm 0.067$ \\
\hline
\end{tabular}
\end{table}

\paragraph{Temporal-window sensitivity}
We next examine whether successful-transition performance depends strongly on
the amount of behavior observed after the target turn. The same samples and
fold assignments are used for all three settings, while the post-turn horizon
is varied between +2\,s, +5\,s, and +10\,s.
Tables~\ref{table:b3_window_auroc} and
\ref{table:b3_window_auprc} report AUROC and AUPRC, respectively.

Performance does not improve consistently as the response horizon becomes
longer. For AUROC, TCN with KP gives the strongest result at +2\,s
($0.572 \pm 0.061$), while TCN with KP+ACC gives the strongest result at
+5\,s ($0.596 \pm 0.043$), and TCN with KP gives the strongest result at
+10\,s ($0.596 \pm 0.074$). The best +5\,s and +10\,s AUROC values differ by
less than 0.001.

A similar pattern is observed for AUPRC. The highest value is obtained by TCN
with KP+ACC at +5\,s ($0.349 \pm 0.063$), followed by TCN with KP at +10\,s
($0.334 \pm 0.086$). Overall, extending the response horizon alone does not
produce a consistent improvement in successful-transition detection.

\begin{table}[ht]
\centering
\caption{AUROC for successful topic-transition detection using different
post-turn observation horizons. Values are the unweighted mean $\pm$ sample
standard deviation across five held-out test folds.}
\label{table:b3_window_auroc}
\begin{tabular}{llccc}
\hline
Model & Input & +2\,s & +5\,s & +10\,s \\
\hline
LR
& KP
& $0.536 \pm 0.031$
& $0.545 \pm 0.025$
& $0.556 \pm 0.026$ \\
& ACC
& $0.516 \pm 0.042$
& $0.525 \pm 0.052$
& $0.546 \pm 0.056$ \\
& KP+ACC
& $0.544 \pm 0.034$
& $0.555 \pm 0.039$
& $0.568 \pm 0.043$ \\
\hline
TCN
& KP
& $\mathbf{0.572 \pm 0.061}$
& $0.549 \pm 0.067$
& $\mathbf{0.596 \pm 0.074}$ \\
& ACC
& $0.503 \pm 0.070$
& $0.488 \pm 0.068$
& $0.483 \pm 0.049$ \\
& KP+ACC
& $0.570 \pm 0.050$
& $\mathbf{0.596 \pm 0.043}$
& $0.565 \pm 0.036$ \\
\hline
TCN--DeepSets
& KP
& $0.566 \pm 0.050$
& $0.563 \pm 0.054$
& $0.539 \pm 0.044$ \\
& ACC
& $0.511 \pm 0.068$
& $0.500 \pm 0.077$
& $0.488 \pm 0.058$ \\
& KP+ACC
& $0.550 \pm 0.032$
& $0.563 \pm 0.067$
& $0.501 \pm 0.091$ \\
\hline
\end{tabular}
\end{table}

\begin{table}[ht]
\centering
\caption{AUPRC for successful topic-transition detection using different
post-turn observation horizons. Values are the unweighted mean $\pm$ sample
standard deviation across five held-out test folds. The positive-class
prevalence is 0.227.}
\label{table:b3_window_auprc}
\begin{tabular}{llccc}
\hline
Model & Input & +2\,s & +5\,s & +10\,s \\
\hline
LR
& KP
& $0.272 \pm 0.044$
& $0.277 \pm 0.040$
& $0.291 \pm 0.029$ \\
& ACC
& $0.264 \pm 0.034$
& $0.267 \pm 0.042$
& $0.274 \pm 0.041$ \\
& KP+ACC
& $0.282 \pm 0.037$
& $0.293 \pm 0.040$
& $0.296 \pm 0.049$ \\
\hline
TCN
& KP
& $\mathbf{0.309 \pm 0.032}$
& $0.275 \pm 0.032$
& $\mathbf{0.334 \pm 0.086}$ \\
& ACC
& $0.229 \pm 0.029$
& $0.242 \pm 0.048$
& $0.218 \pm 0.029$ \\
& KP+ACC
& $0.284 \pm 0.033$
& $\mathbf{0.349 \pm 0.063}$
& $0.286 \pm 0.058$ \\
\hline
TCN--DeepSets
& KP
& $0.287 \pm 0.063$
& $0.295 \pm 0.028$
& $0.274 \pm 0.072$ \\
& ACC
& $0.232 \pm 0.041$
& $0.252 \pm 0.054$
& $0.235 \pm 0.031$ \\
& KP+ACC
& $0.276 \pm 0.016$
& $0.336 \pm 0.097$
& $0.279 \pm 0.067$ \\
\hline
\end{tabular}
\end{table}

\paragraph{Secondary attempt-detection task}
As a secondary analysis, we evaluate whether nonverbal behavior can distinguish
topic-redirection attempts from ordinary conversation, irrespective of whether
the attempt subsequently becomes established. For this task, the 316 successful
and 148 failed attempts are combined into a positive class of 464 samples,
while the 928 ordinary conversational controls form the negative class. The
positive prevalence is therefore 33.3\%.

The secondary task uses the same samples, fold assignments, models,
modalities, and temporal-window settings as the main benchmark.
Tables~\ref{table:b3_attempt_auroc} and
\ref{table:b3_attempt_auprc} report the corresponding AUROC and AUPRC results.

Attempt-detection performance changes only modestly across the three temporal
horizons. LR with ACC gives the highest AUROC at all three horizons:
$0.557 \pm 0.026$ at +2\,s, $0.568 \pm 0.029$ at +5\,s, and
$0.563 \pm 0.025$ at +10\,s. The highest AUPRC is
$0.411 \pm 0.054$, obtained by LR with KP+ACC at +10\,s. These results show
little evidence of a systematic improvement as more post-turn behavior is
included.

\begin{table}[ht]
\centering
\caption{AUROC for the secondary topic-redirection attempt-detection task
using different post-turn observation horizons. Values are the unweighted
mean $\pm$ sample standard deviation across five held-out test folds.}
\label{table:b3_attempt_auroc}
\begin{tabular}{llccc}
\hline
Model & Input & +2\,s & +5\,s & +10\,s \\
\hline
LR
& KP
& $0.524 \pm 0.073$
& $0.528 \pm 0.073$
& $0.534 \pm 0.037$ \\
& ACC
& $\mathbf{0.557 \pm 0.026}$
& $\mathbf{0.568 \pm 0.029}$
& $\mathbf{0.563 \pm 0.025}$ \\
& KP+ACC
& $0.547 \pm 0.071$
& $0.544 \pm 0.057$
& $0.552 \pm 0.032$ \\
\hline
TCN
& KP
& $0.519 \pm 0.020$
& $0.528 \pm 0.041$
& $0.520 \pm 0.027$ \\
& ACC
& $0.504 \pm 0.051$
& $0.508 \pm 0.082$
& $0.502 \pm 0.078$ \\
& KP+ACC
& $0.541 \pm 0.031$
& $0.538 \pm 0.042$
& $0.549 \pm 0.079$ \\
\hline
TCN--DeepSets
& KP
& $0.498 \pm 0.055$
& $0.519 \pm 0.047$
& $0.509 \pm 0.031$ \\
& ACC
& $0.502 \pm 0.064$
& $0.512 \pm 0.057$
& $0.541 \pm 0.049$ \\
& KP+ACC
& $0.518 \pm 0.050$
& $0.506 \pm 0.061$
& $0.529 \pm 0.057$ \\
\hline
\end{tabular}
\end{table}

\begin{table}[ht]
\centering
\caption{AUPRC for the secondary topic-redirection attempt-detection task
using different post-turn observation horizons. Values are the unweighted
mean $\pm$ sample standard deviation across five held-out test folds. The
positive-class prevalence is 0.333.}
\label{table:b3_attempt_auprc}
\begin{tabular}{llccc}
\hline
Model & Input & +2\,s & +5\,s & +10\,s \\
\hline
LR
& KP
& $0.382 \pm 0.066$
& $0.383 \pm 0.072$
& $0.387 \pm 0.056$ \\
& ACC
& $0.384 \pm 0.034$
& $\mathbf{0.401 \pm 0.040}$
& $0.398 \pm 0.034$ \\
& KP+ACC
& $\mathbf{0.395 \pm 0.065}$
& $0.399 \pm 0.062$
& $\mathbf{0.411 \pm 0.054}$ \\
\hline
TCN
& KP
& $0.357 \pm 0.023$
& $0.380 \pm 0.046$
& $0.373 \pm 0.035$ \\
& ACC
& $0.354 \pm 0.029$
& $0.357 \pm 0.053$
& $0.342 \pm 0.039$ \\
& KP+ACC
& $0.393 \pm 0.030$
& $0.383 \pm 0.040$
& $0.392 \pm 0.047$ \\
\hline
TCN--DeepSets
& KP
& $0.327 \pm 0.026$
& $0.373 \pm 0.070$
& $0.356 \pm 0.036$ \\
& ACC
& $0.350 \pm 0.053$
& $0.354 \pm 0.032$
& $0.370 \pm 0.032$ \\
& KP+ACC
& $0.366 \pm 0.044$
& $0.348 \pm 0.048$
& $0.376 \pm 0.042$ \\
\hline
\end{tabular}
\end{table}

\paragraph{Summary}
Across the three response horizons, successful topic-transition detection
shows modest discrimination, with several configurations performing above the
AUPRC class-prevalence baseline. Increasing the post-turn observation horizon
from 2\,s to 10\,s does not lead to a consistent improvement. The strongest
successful-transition results are obtained by initiator-level TCN models,
while the tested TCN--DeepSets baseline does not provide a consistent
improvement from incorporating the surrounding group.

The secondary attempt-detection task is less sensitive to temporal-window
length. Its strongest results are obtained by the LR baselines, particularly
with acceleration. Overall, the results indicate that nonverbal behavior
contains some information about both topic-redirection attempts and successful
topic transitions, but the baseline performance remains modest.

\paragraph{Running details and compute resources}
All Benchmark 3 experiments were executed on the internal Slurm cluster.
Neural-network experiments used one GPU, four CPU cores, and 32\,GB of system
memory per job, with a maximum wall time of 12 hours. Logistic-regression
experiments were executed on CPU using four CPU cores and 16\,GB of memory,
with a maximum wall time of 4 hours.